\documentclass[preprintnumbers, floatfix, onecolumn, superscriptaddress,nofootinbib,10pt]{revtex4-2}

\usepackage{graphicx}
\usepackage{epsfig}
\usepackage{bm}
\usepackage{amssymb}
\usepackage{float}
\usepackage{amsmath}
\usepackage{dcolumn}
\usepackage{cancel}

\usepackage{mathrsfs}
\usepackage{dcolumn}
\usepackage{graphicx}
\usepackage{amsmath}
\usepackage{amsfonts}
\usepackage{amssymb}
\usepackage{microtype}
\usepackage{subfigure}
\usepackage{makeidx}
\usepackage{bm}
\usepackage{epsf}
\usepackage{color}
\usepackage{multirow,dcolumn}
\usepackage{graphicx}
\usepackage{mathrsfs}
\graphicspath{{Images/}}

\def\doi{http://doi.org}

\def\be{\begin{equation*}}
\def\ee{\end{equation*}}

\begin{document}

\title{Black Holes with Scalar Hair in Three Dimensions}

\author{Thanasis Karakasis}
\email{thanasiskarakasis@mail.ntua.gr}
\affiliation{Physics Department, National Technical University of Athens, 15780 Zografou Campus, Athens, Greece}
	
\author{George Koutsoumbas}
	\email{kutsubas@central.ntua.gr}
	\affiliation{Physics Department, National Technical University of Athens, 15780 Zografou Campus, Athens, Greece}

\author{Eleftherios Papantonopoulos}
\email{lpapa@central.ntua.gr} \affiliation{Physics Department, National Technical University of Athens, 15780 Zografou Campus, Athens, Greece}

\begin{abstract}
Three-dimensional static and spinning black hole solutions of the Einstein-Klein-Gordon system are obtained for a particular scalar field configuration. At large distances, and for small scalar field, the solutions reduce to the BTZ black hole. The scalar field dresses the black hole with secondary scalar hair, since the scalar charge is related to the conserved black hole mass. A self interacting potential is included, containing a mass term that is above the Breitenlohner-Freedman bound in three dimensions. Independence of the scalar potential from the conserved black hole charges, imposes fixed mass and angular momentum to scalar charge ratios. The thermodynamic properties as well as the energy conditions of the black hole are analysed.
\end{abstract}
\maketitle

\tableofcontents

\section{Introduction}

Gravity in three dimensions has been studied extensively and it is interesting in many ways. One of the main advantages of working
in three dimensions is that space-time has simple enough topology and it  can be characterized completely and explicitly, giving  important insights into black hole physics and the structure of quantum gravity and it also can provide information on how the dimensionality of space-time affects the resulting physics. Einstein's theory of gravitation, General Relativity (GR) in four or higher dimensions and in the absence of matter, admits Schwarzchild black holes, while  in three dimensions the Weyl tensor, which describes the distortion of the shape of a body in the presence of the  gravitational force, vanishes by definition while the Ricci tensor, describing  how this force changes the volume of the body, vanishes in the absence of matter. Therefore, since $Riemann = Weyl + Ricci$ we can only have flat space-time. Hence the theory can only admit a conical singularity as a solution of the field equations \cite{Deser:1983tn}. As a result, a negative cosmological constant was included to make the Ricci tensor non-zero and the BTZ black hole has been found \cite{Banados:1992wn,Banados:1992gq} by Ba\~{n}ados, Teitelboim, and Zanelli. The BTZ black hole possesses two black hole horizons, it can arise from collapsing matter \cite{Carlip:1995qv}, and it has well posed thermodynamic properties: it quantizes, for example, the entropy is given by a Bekenstein-Hawking area law, where the area is replaced by the circumference of the black hole. The $fundamental$ length scale of the three-dimensional black hole is the AdS radius, since the mass term turns out to be a dimensionless constant.

After the discovery of the BTZ black hole, scalar fields minimally and non-minimally coupled to gravity were introduced as matter fields. In \cite{Martinez:1996gn,Henneaux:2002wm} three-dimensional black holes with a conformally coupled scalar field, being regular everywhere, were discussed. The entropy in these cases is modified due to the non-minimal coupling between matter and gravity. After these first results other hairy black holes in three dimensions were discussed  \cite{CMT1,CMT2,Correa:2012rc,Natsuume,Aparicio:2012yq,Xu:2014uha}. In \cite{Cardenas:2014kaa}  three-dimensional gravity with negative cosmological constant in the presence of a scalar field and an Abelian gauge field was introduced. Both fields are conformally coupled to gravity, the scalar field through a non-minimal coupling with the curvature and the gauge field by means of a Lagrangian given by a power of the Maxwell field. A sixth-power self-interaction potential, which does not spoil conformal invariance is also included in the action, resulting to a very simple relation between the scalar curvature and the cosmological constant. The cases where the entropy is positive were pointed out, the authors arguing that the appearance of non-physical negative entropy in the Jordan frame results in a naked singularity in the Einstein frame.

$(2+1)$-dimensional charged black holes with scalar hair where derived, where the scalar potential is not fixed ad hoc, but derived from Einstein's equations in \cite{Xu:2014uka} and \cite{Xu:2013nia}. In \cite{Tang:2019jkn} exact three-dimensional black holes with a non-minimal scalar field were discussed, the coupling parameter being left arbitrary. In \cite{Becar:2019hwk,Gonzalez:2019xfr} the motion and trajectories of photons and the quasi-normal modes of three-dimensional rotating Ho\v{r}ava-AdS black hole were calculated.
In \cite{Chan:1994qa, Chan:1995wj, Chan:1996rd},  static black holes in three-dimensional dilaton gravity and modifications of the BTZ black hole by a dilaton scalar were investigated, the dilaton case attracting more attention over the years \cite{Yamazaki:2001ue, HosseinHendi:2017soi, Dehghani:2017thu, powermax, Dehghani:2017zkm}.
Three-dimensional black holes in modified theories of gravity have also been studied.  $f(R)$ gravity black holes are found in \cite{Nashed:2021jvn, Tang:2019qiy, Hendi:2014wsa, Hendi:2014mba}, while BTZ-like black holes in the Einstein Gauss Bonnet theory have been recently obtained \cite{Konoplya:2020ibi,Ma:2020ufk,Hennigar:2020fkv,Hennigar:2020drx,Jusufi:2023fpo}, while scalar fields in three-dimensional modified gravity have also been investigated in the Horndeski theory \cite{Bravo-Gaete:2014haa} also in a general scalar tensor theory as well \cite{Baake:2020tgk}. Finally, scalar fields have been added to three-dimensional $f(R)$ gravity and the resulting black holes were discussed \cite{Karakasis:2021lnq, Karakasis:2021ttn}.

As we already discussed, a scale has to be introduced  with the presence of a cosmological constant, to generate the BTZ black hole in $(2+1)$-dimensions. In this work we will follow another approach. We will introduce a scalar field, parametrizing a matter distribution, which back-reacts on a $(2+1)$-dimensional metric. The scalar field is characterized by a parameter called scalar charge. When the scalar field back - reacts to the space-time metric, this scalar charge appears in the metric introducing an effective cosmological constant. Then a hairy black hole is generated which at small distances because matter is strongly back - reacting to the metric a new BTZ-like black hole is generated, while at large distances because the back - reaction is weak, the BTZ black hole is found. We  discuss the thermodynamic properties and the energy conditions of the hairy BTZ-like black hole solution as well as the rotating case.

This work is organized as follows. In section \ref{sec2} we derive static solutions of the Einstein-Klein-Gordon system and discuss the geometric properties of the new black hole solutions, as well as the energy conditions and the thermodynamic properties of these solutions. In section \ref{sec3} we derive novel rotating solutions that reduce to the ones obtained in section \ref{sec2}, once the angular momentum of the black hole vanishes and discuss their thermodynamics. We discuss the resultant scalar potentials, which from a field theory point-of-view have to be independent of the black hole parameters, a fact implying that the black holes produced have a fixed mass to angular momentum ratio. Finally, in \ref{sec4} we point out the prospects of our work.

\section{Static Black Hole Solutions}\label{sec2}

We consider a simple model of a scalar field minimally coupled to gravity in three dimensions in the action
\begin{equation} S = \frac{1}{8\pi}\int d^3x\sqrt{-g}\left\{\frac{R}{2}  - \frac{1}{2}\partial^{\alpha}\phi\partial_{\alpha}\phi - V(\phi)\right\}~, \label{theory} \end{equation}
which consists of the Ricci scalar and a self interacting scalar field minimally coupled to gravity. By variation of this action we obtain the field equations
\begin{eqnarray}
&&G_{\mu\nu} \equiv R_{\mu\nu}-\frac{1}{2}g_{\mu\nu}R  = T_{\mu\nu}~, \label{einstein} \\
&&\Box\phi - \frac{\partial V}{\partial \phi}=0~, \label{box} \\
&&T_{\mu\nu} = \partial_{\mu}\phi\partial_{\nu}\phi-\frac{1}{2}g_{\mu\nu}\partial^{\alpha}\phi\partial_{\alpha}\phi - g_{\mu\nu}V(\phi)~.
\end{eqnarray}
To solve the field equations we assume that the scalar field has a coulomb-like form as in the four dimensional theory \cite{Bakopoulos:2021dry}, with the solution first appearing in \cite{Herdeiro:2015waa}. We find that the scalar field dresses the black hole with secondary hair, with the scalar charge appearing in the conserved black hole mass. The Null Energy Condition is violated inside the event horizon, a feature also appearing in the four-dimensional sibling \cite{Bakopoulos:2021dry}, and we point out that this is a global feature in any space-time having a vanishing $1/g_{rr}(r_+)=0$ at the event horizon of the black hole, where only radial dependence on the scalar field is assumed. At large distances, the solution reduces to the BTZ black hole, while the scalar potential admits an even power law expansion, with the mass term being above the Breitenlohner-Freedman bound.

We consider a $(2+1)$-dimensional metric ansatz of the form
\begin{equation} ds^2 = -h(r)dt^2 +\frac{1}{b(r)}dr^2 +r^2d\theta^2~,\end{equation}
where $h(r), b(r)$ are the two unknown metric functions to be found  solving the field equations. Since the metric only depends on $r,$ we will consider that $\phi$ is only $r$-dependent, hence $\phi=\phi(r), V=V(r)$. The $tt,rr,\theta\theta$ components of the Einstein field equation and the Klein-Gordon equation read
\begin{eqnarray}
&&b'(r)+r b(r) \phi '(r)^2+2 r V(r) =0~,\\
&&b(r) \left(\frac{h'(r)}{r h(r)}-\phi '(r)^2\right)+2 V(r) =0~,\\
&&-b(r) h'(r)^2+2 h(r)^2 \left(b(r) \phi '(r)^2+2 V(r)\right)+h(r)\left(b'(r) h'(r)+2 b(r) h''(r)\right)=0~,\\
&&\frac{1}{2} b'(r) \phi '(r)+b(r) \left(\left(\frac{h'(r)}{2 h(r)}+\frac{1}{r}\right) \phi '(r)+\phi ''(r)\right)-\frac{V'(r)}{\phi '(r)}=0~. \label{KG1}
\end{eqnarray}
Using the Bianchi identity one can prove that (\ref{KG1}) can be obtained from the Einstein equations. Hence, we have a system of with three independent equations in four unknown functions. As a result one of the unknowns has to be fixed $ad~hoc$. However, if one fixes the scalar potential from the begining in (\ref{theory}), then the system can be, at least in principle, integrated, since one will have three unknown functions and three equations.
Hence, we fix the form of the scalar field as
\begin{equation}\phi(r) = \frac{A}{r}~, \label{phi} \end{equation}
where $A$ is a scalar length scale that controls the behaviour of the scalar field, which we will call scalar charge. Note that the scalar charge has been defined as the term controlling the $\mathcal{O}(r^{-1})$ term of the scalar field at infinity in four dimensions and  we will use the same terminology in our work. We have checked that using $\phi(r)= A/r^n$ with $n>0,$ exact results can be obtained for different $n,$ however for $n=1$ we can find simple exact solutions.

\begin{figure}
\centering
\includegraphics[width=.40\textwidth]{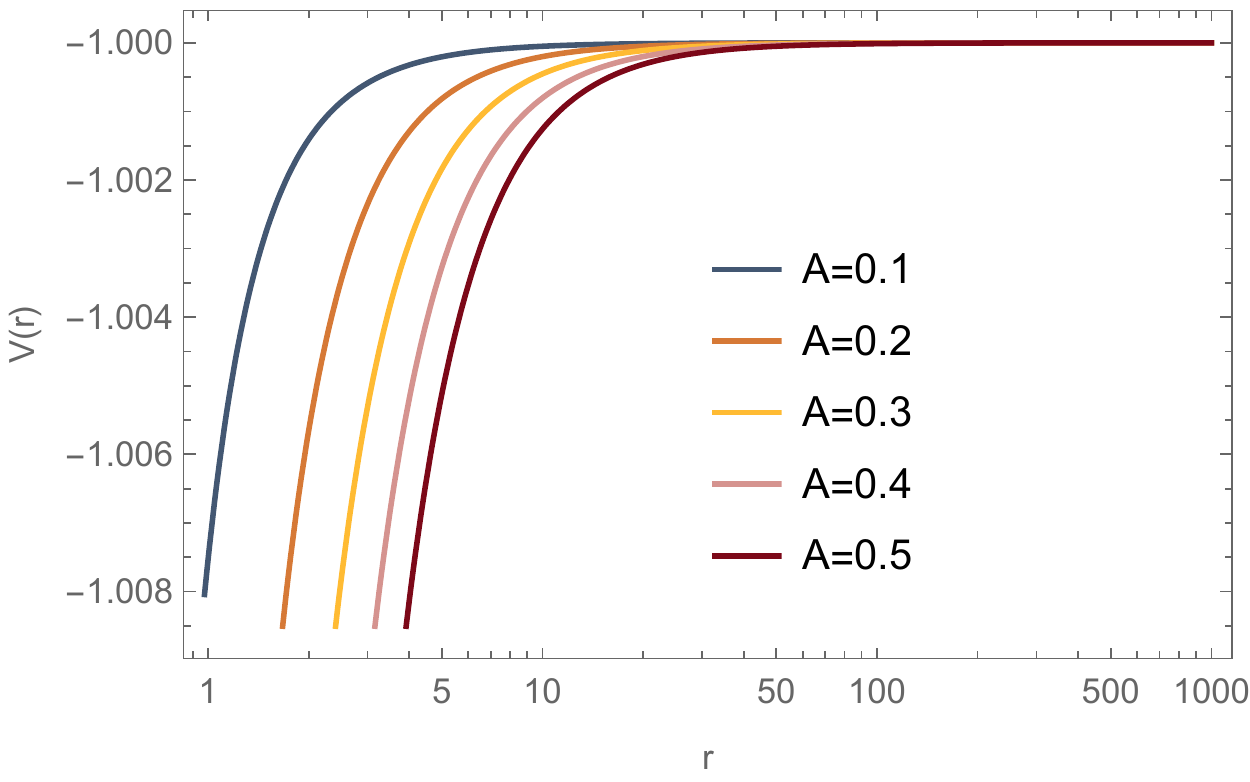}
\includegraphics[width=.40\textwidth]{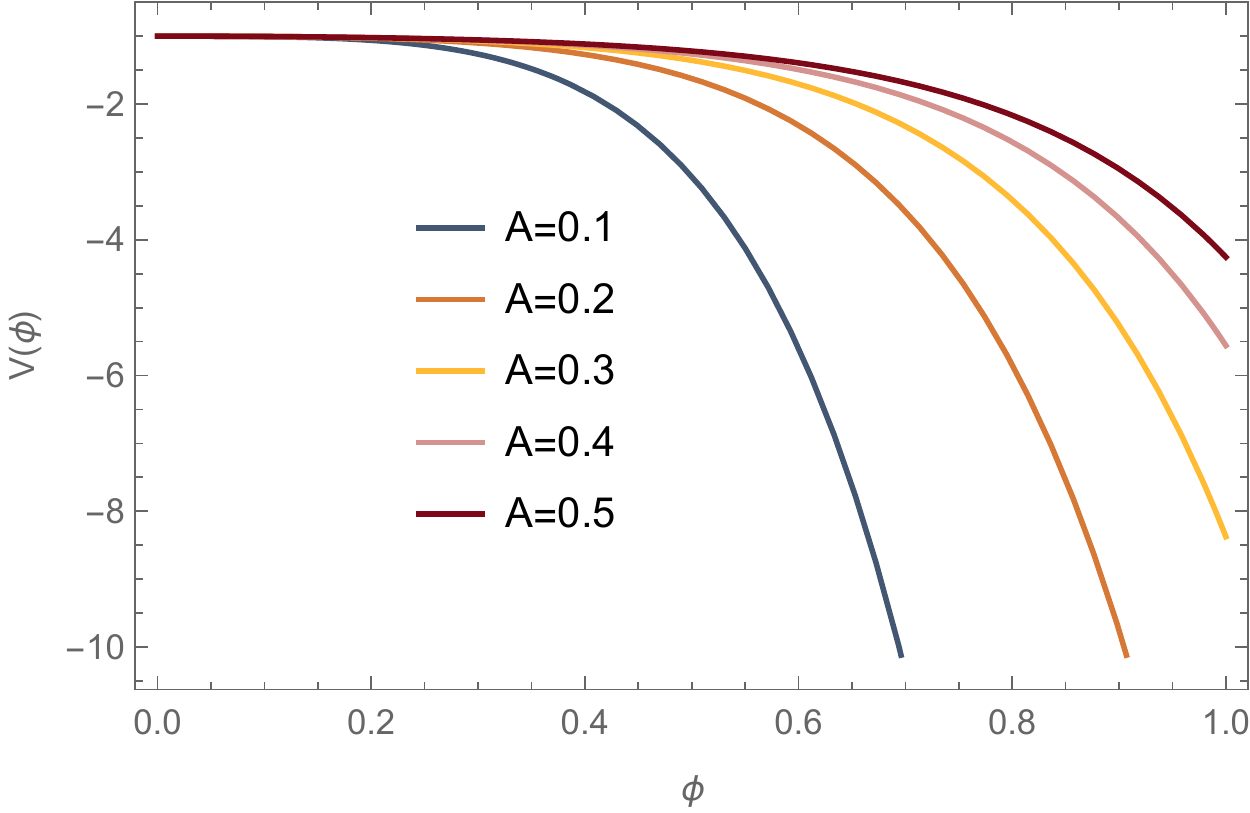}
\caption{The potential $V(r)$ and $V(\phi)$ for $\mathcal{M}=-\Lambda=1$, while changing the scalar charge $A$.} \label{V}
\end{figure}

 There is a pole in the scalar field function for $r\to 0 $, however this will not be a problem because of the presence of a curvature singularity at $r=0$ as we will discuss, thence the solution will be valid for $r>0$. Then, we find that
\begin{eqnarray}
&&b(r) = \frac{c_1 r^2 }{A^2 c_3^2}e^{\frac{A^2}{2 r^2}} \left(A^2 c_2 e^{\frac{A^2}{2 r^2}}+c_3\right)~,\\
&&h(r) = r^2 \left(\frac{c_3}{A^2} e^{-\frac{A^2}{2 r^2}}+c_2\right)~,\\
&&V(r) = \frac{c_1 c_2 }{2 c_3^2 r^2}e^{\frac{A^2}{r^2}} \left(A^2-2 r^2\right)-\frac{c_1 }{A^2 c_3}e^{\frac{A^2}{2 r^2}}~,
\end{eqnarray}
where $c_1,c_2,c_3$ are constants of integration to be determined from the boundary conditions. Note here that at large distances, the solution asymptotes to
\begin{eqnarray}
&&b(r\to\infty) \sim \frac{c_1 r^2 }{c_3^2}\left(\frac{c_3}{A^2}+c_2\right)+\frac{c_1 \left(2 A^2 c_2+c_3\right)}{2 c_3^2}+\frac{A^2 c_1 \left(4 A^2 c_2+c_3\right)}{8 c_3^2 r^2}+\mathcal{O}\left(\left(\frac{1}{r}\right)^4\right)~,\\
&&h(r\to\infty) \sim +\frac{A^2 c_3}{8 r^2}+\frac{c_3 r^2}{A^2}+c_2 r^2-\frac{c_3}{2} +\mathcal{O}\left(\left(\frac{1}{r}\right)^4\right)~,\\
&&V(r\to\infty) \sim -\frac{c_1 }{c_3^2}\left(\frac{c_3}{A^2}+c_2\right)-\frac{A^2 c_1 c_2+c_1 c_3}{2 c_3^2 r^2}+\mathcal{O}\left(\left(\frac{1}{r}\right)^4\right)~.
\end{eqnarray}
From the above relations we can see that we have some $\mathcal{O}(r^2)$ terms that survive at large distances which we can identify as cosmological constant terms, that are generated by the vacuum of the scalar field theory and depend on $A$. For $c_1=A^4 (c_2+\Lambda )^2, c_3 = -A^2 (c_2+\Lambda )$, where $\Lambda$ is an effective cosmological constant, we rewrite the asymptotic expressions of the metric functions at large distances
\begin{eqnarray}
&&b(r\to\infty) \sim -\Lambda  r^2+\frac{1}{2} A^2 (c_2-\Lambda )+\frac{A^4 (3 c_2-\Lambda )}{8 r^2}+\mathcal{O}\left(\left(\frac{1}{r}\right)^4\right)~,\\
&&h(r\to\infty) \sim -\Lambda  r^2+\frac{1}{2} A^2 (c_2+\Lambda )-\frac{A^4 (c_2+\Lambda )}{8 r^2}+\mathcal{O}\left(\left(\frac{1}{r}\right)^4\right)~.
\end{eqnarray}
To compute the mass of the black hole, we will use the quasi-local method \cite{Chan:1995wj}. The quasi-local energy at a finite distance $r_0$ is defined as
\begin{equation} E(r_0) = 2\left(\sqrt{b_0(r_0)}-\sqrt{b(r_0)}\right)~,\end{equation}
where $b_0$ determines the zero of the energy (which we take to be the pure AdS space-time $b_0(r)=-\Lambda r^2$)  and the quasi-local mass at $r_0$ can be obtained as
\begin{equation} m(r_0) = \sqrt{h(r_0)}E(r_0)~.\end{equation}
Now, the ADM mass of the black hole can be read off by taking the limit at $r_0\to\infty$
\begin{equation} \mathcal{M} = -\frac{1}{2}A^2\left(c_2-\Lambda\right)~. \label{mass} \end{equation}
Setting $c_2 = \cfrac{A^2 \Lambda -2 \mathcal{M}}{A^2}$ we can rewrite the solution in terms of the black hole mass $\mathcal{M}$, the scalar charge $A$ and the effective cosmological constant $\Lambda$, which are the parameters of our solution
\begin{eqnarray}
&&h(r) = r^2 \left(\frac{A^2 \Lambda -2 \mathcal{M}}{A^2}-e^{-\frac{A^2}{2 r^2}} \left(\frac{A^2 \Lambda -2
   \mathcal{M}}{A^2}+\Lambda \right)\right),\\
&&b(r) = \frac{2 r^2 e^{\frac{A^2}{2 r^2}} \left( \mathcal{M}-A^2 \Lambda \right)-r^2 e^{\frac{A^2}{r^2}}
   \left(2 \mathcal{M}-A^2 \Lambda \right)}{A^2}~,\\
&&V(r) =  \frac{e^{\frac{A^2}{r^2}} \left(A^2-2 r^2\right) \left(A^2 \Lambda -2 \mathcal{M}\right)}{2 A^2
   r^2}+\frac{2 e^{\frac{A^2}{2 r^2}} \left(A^2 \Lambda - \mathcal{M}\right)}{A^2},\\
&&V(\phi) = \frac{2 e^{\frac{\phi ^2}{2}} \left(A^2 \Lambda - \mathcal{M}\right)}{A^2}+\frac{e^{\phi ^2}
   \left(\phi ^2-2\right) \left(A^2 \Lambda -2 \mathcal{M}\right)}{2 A^2}~ \label{potential}.
\end{eqnarray}
Their asymptotic expressions now read
\begin{eqnarray}
&&h(r\to\infty) \sim -\Lambda  r^2+\left(A^2 \Lambda -\mathcal{M}\right)-\frac{A^2 \left(A^2 \Lambda -\mathcal{M}\right)}{4
   r^2}+\mathcal{O}\left(\left(\frac{1}{r}\right)^4\right)~,\\
&&b(r\to\infty) \sim -\Lambda  r^2-\mathcal{M}+\frac{A^4 \Lambda -3 A^2 \mathcal{M}}{4 r^2}+\mathcal{O}\left(\left(\frac{1}{r}\right)^4\right)~,\\
&&V(r\to\infty) \sim \Lambda + \frac{A^2 \Lambda }{2 r^2}+\frac{A^2 \left(A^2 \Lambda -\mathcal{M}\right)}{4 r^4}+\frac{3 A^6 \Lambda -5 A^4 \mathcal{M}}{24 r^6}+\mathcal{O}\left(\left(\frac{1}{r}\right)^8\right)~, \label{pt}\\
&&V(\phi \to 0) \sim \Lambda +\frac{\Lambda  \phi ^2}{2}+\phi ^4 \left(\frac{\Lambda
   }{4}-\frac{ \mathcal{M}}{4A^2}\right) + \phi ^6 \left(\frac{\Lambda }{8}-\frac{5 \mathcal{M}}{3 A^2}\right) + \mathcal{O}\left(\phi ^{8}\right)~.
\end{eqnarray}
We can see that at large distances, the solution resembles the BTZ black hole, while corrections in the structure of space-time appear as $\mathcal{O}\left(\left(\cfrac{1}{r}\right)^s\right)$ terms, where $s\geq2$ which are supported by the existence of the scalar field. Moreover, the scalar field dresses the black hole with secondary scalar hair, since the conserved mass is given by the scalar charge in addition to an integration constant. The potential has a mass term given by $m^2=V''(\phi=0) = \Lambda,$ which is above the Breitenlohner-Freedman bound in three dimensions \cite{Breitenlohner:1982jf,Mezincescu:1984ev} and for small $\phi$ (large $r$), the potential admits an even power series expansion. It is also invariant under the substitution $\phi\to -\phi$. The potential (\ref{potential}) contains both the mass and the scalar charge of the black hole space-time. However this should not happen. As a result we have to find a way to render the potential independent of the black hole mass and scalar charge. By inspection we can see that we can define the conserved mass to scalar charge ratio $q$ as
\begin{equation} \label{cond} q = \frac{\mathcal{M}}{A^2}~,\end{equation}
and now the potential will be
\begin{eqnarray}
&&V(\phi) =
e^{\frac{\phi ^2}{2}} (2 \Lambda -2 q)+e^{\phi ^2} \left(\frac{\Lambda  \phi ^2}{2}-\Lambda -q \phi ^2+2 q\right)~,\\
&&V(\phi\to0) \sim \Lambda +\frac{\Lambda  \phi ^2}{2}+ \frac{\Lambda -q}{4}\phi ^4+
   \frac{3 \Lambda -5 q}{24} \phi ^6+\mathcal{O}\left(\phi ^8\right)
\end{eqnarray}
where now $q$ is the parameter of our theory. Now the potential is general enough and the theory (\ref{theory}) can yield black holes with different masses and scalar charges. The mass of the resulting compact object might reduce through Hawking evaporation for example, but a varying mass implies a varying scalar charge, so that their ratio $q$ is constant. Therefore, from a field theory point-of-view we can argue that the scalar charge $A$ is kind of a thermodynamic variable, since it has to vary when $\mathcal{M}$ is changing. In Fig.~\ref{V} we plot the scalar potential as a function of $r$ and $\phi$, where we can see that the potential is always negative in order to support the hairy structure and to violate the no-hair theorem. Moreover from the plot of $V(\phi)$ it is clear that the theory contains a global maximum located at $V_{\text{max}}(\phi) = \Lambda$.
We have also checked that the on-shell action is constant at large distances.
The $1/g_{rr}$ component has two roots given by
\begin{equation} r_{\pm} = \pm A\left(\ln \left(\frac{4 \left(A^2 \Lambda - \mathcal{M}\right)^2}{\left(A^2 \Lambda -2
   \mathcal{M}\right)^2}\right)\right)^{-1/2}~.\end{equation}
We work on the solution in AdS space-time, the horizon being $r_+$. From now on, we will set $\Lambda=-\ell^{-2}$, where $\ell$ denotes the AdS radius. In addition, note that there always exist a horizon when $\mathcal{M}$ and $A$ are positive and the scalar field does not imply any bound for the existence of a horizon.
In the limit of small scalar charge $A$, we obtain, at zero order, the BTZ black hole
\begin{eqnarray}
&&h(r) \sim r^2 \left(\frac{1}{\ell ^2}-\frac{\mathcal{M}}{r^2}\right)+A^2 r^2
   \left(\frac{\mathcal{M}}{ 4r^4}-\frac{1}{r^2 \ell ^2}\right)+\mathcal{O}\left(A^4\right)~,\\
&&b(r) \sim r^2 \left(\frac{1}{\ell ^2}-\frac{\mathcal{M}}{r^2}\right)-\frac{3 A^2 \mathcal{M}}{
   4r^2}+\mathcal{O}\left(A^4\right)~,\\
&&V(r) \sim -\frac{1}{\ell^2}-\frac{A^2 \left(2 r^2+\ell ^2 \mathcal{M}\right)}{4 \left(r^4 \ell
   ^2\right)}-\frac{A^4 \left(3 r^2+20 \ell ^2 \mathcal{M}\right)}{12 \left(r^6 \ell
   ^2\right)}+\mathcal{O}\left(A^6\right)~.
\end{eqnarray}
This is related to the asymptotic nature of the scalar field. In our case, the scalar field decays fast enough ($\mathcal{O}(r^{-1})$) and hence its impact on the conserved mass is mild, in the sense that the mass is not completely determined by the scalar hair parameter $A$. As a result, sending $\phi$ to zero we can obtain the massive BTZ black hole. We note that this is not the case when $\phi$ falls like $\mathcal{O}(r^{-1/2})$ \cite{Martinez:1996gn, Henneaux:2002wm, CMT1, Correa:2012rc, Xu:2014uha, Cardenas:2014kaa, Xu:2014uka} where the mass is given explicitly in terms of the scalar hair parameter. In our case, the mass (\ref{mass}) depends on the scalar hair $A$ but also on an independent integration constant $c_2$. Therefore, the integration constant can take any particular value and we can always have a massive black hole solution when $A$ approaches zero and the scalar field vanishes. Recently another solution appeared \cite{Desa:2022gtw}, where the scalar field falls faster at infinity ($\mathcal{O}(r^{-1})$ as in our case) and the no-hair limit in this case is also well-defined and the BTZ black hole is obtained in the limit of vanishing scalar field.

\begin{figure}[h]
\centering
\includegraphics[width=.40\textwidth]{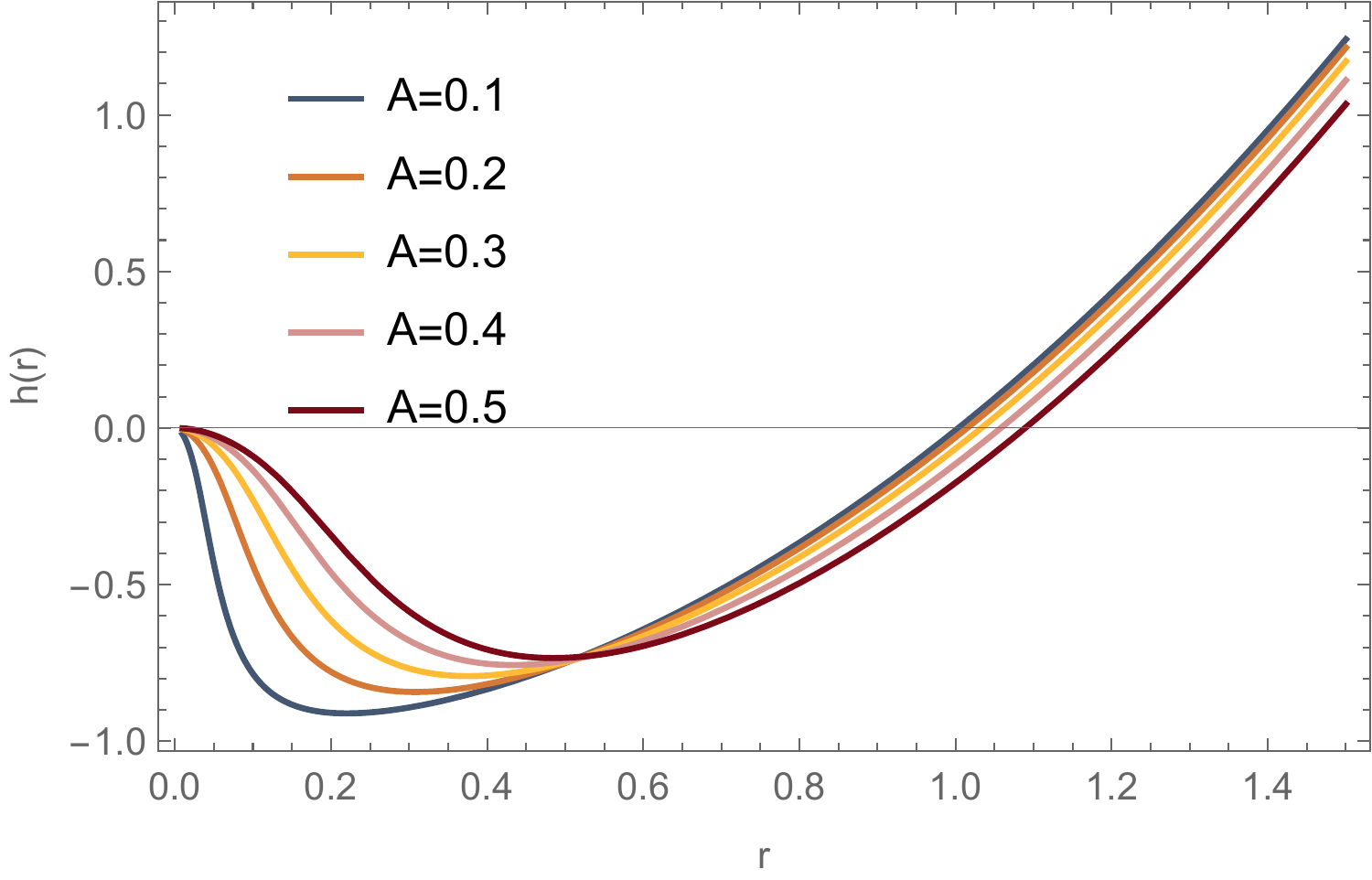}
\includegraphics[width=.38\textwidth]{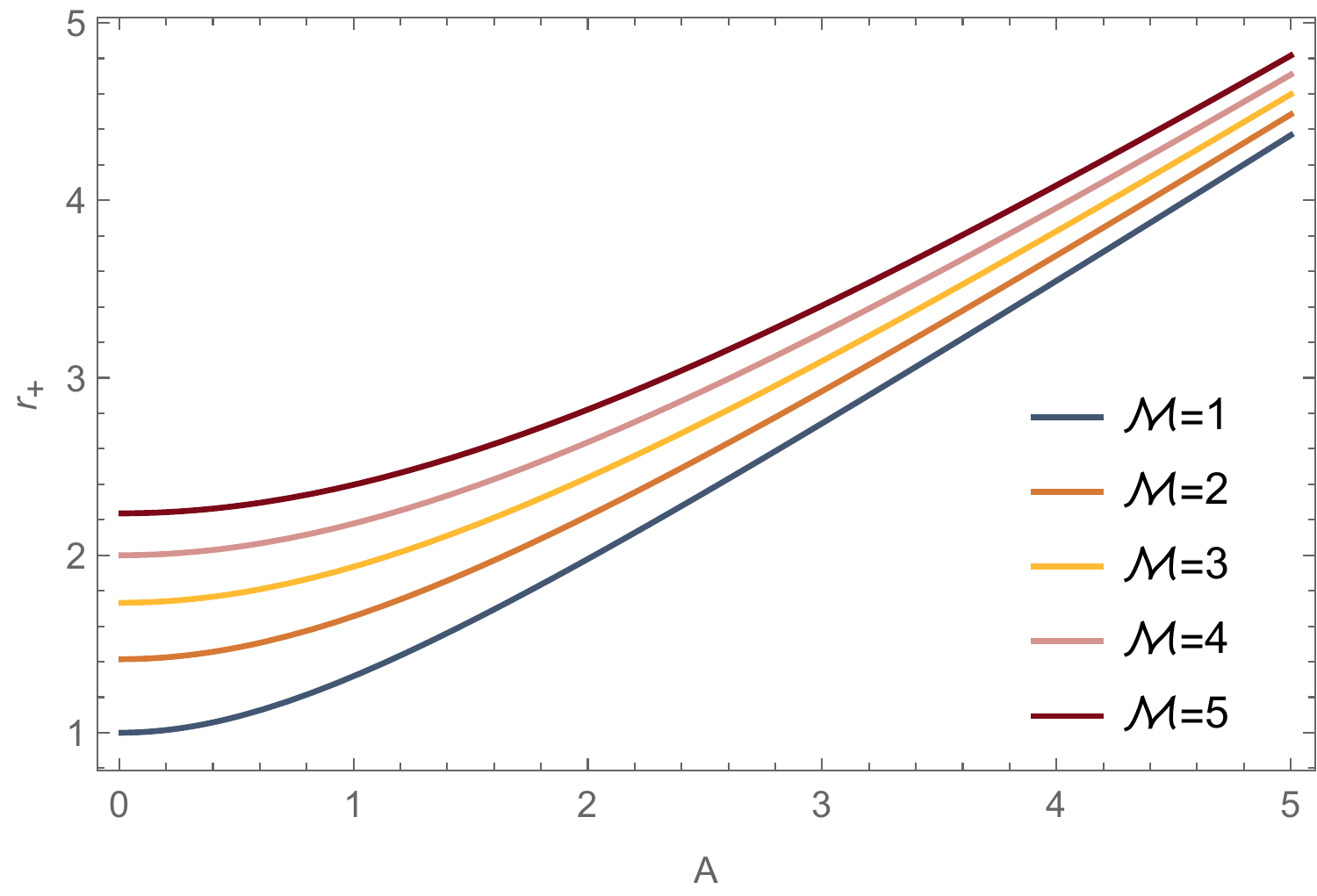}
\caption{Left: $h(r)$ versus $r$ for $\mathcal{M}=\ell=1$, while changing the scalar charge $A$. Right: $r_+$ as a function of $A$, while changing the mass of the black hole for $\ell=1$.} \label{horizon}
\end{figure}
\begin{figure}[h]
\centering
\includegraphics[width=.40\textwidth]{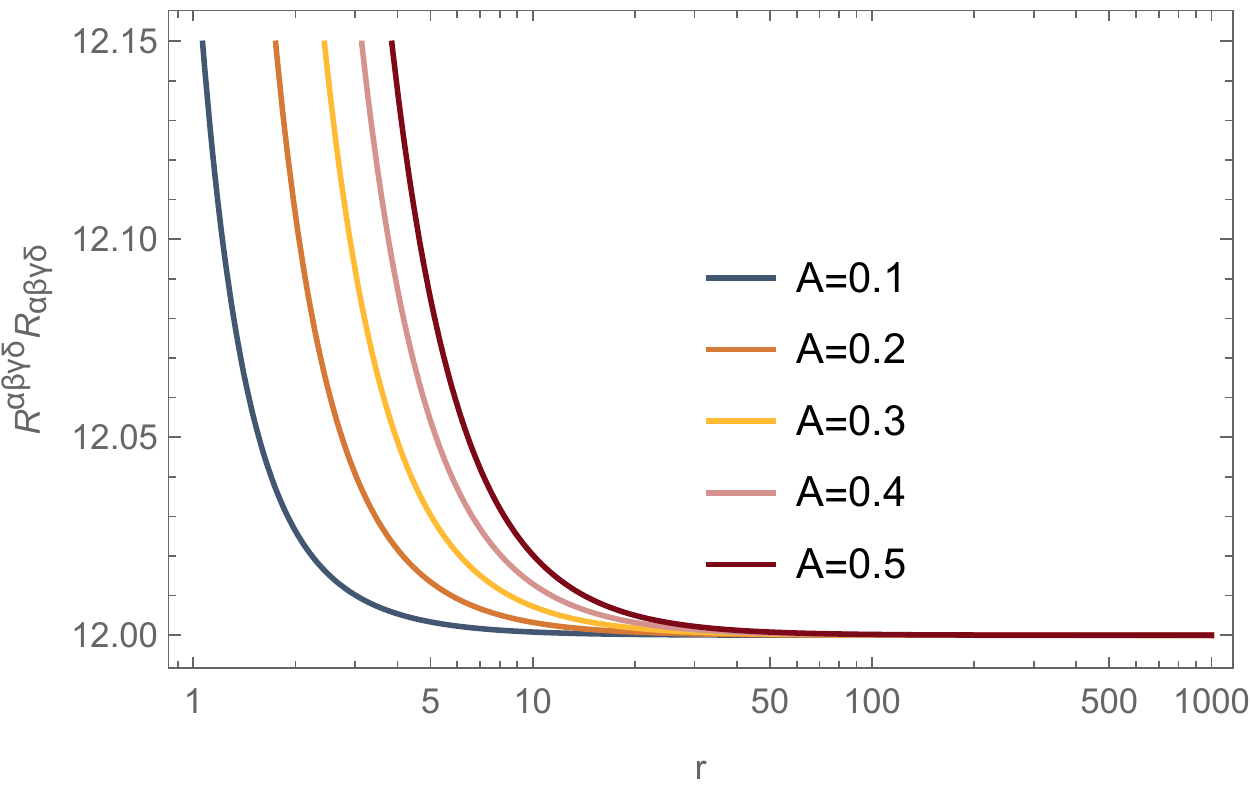}
\caption{The Kretschmann scalar $R_{\alpha\beta\gamma\delta}R^{\alpha\beta\gamma\delta}(r)$ for $\mathcal{M}=-\Lambda =1/\ell^2=1$ while changing $A$.} \label{kre}
\end{figure}

In Fig.~\ref{horizon} we plot $h(r)$ as a function of $r$ for different scalar charges and $r_+$ as a function of the scalar charge $A$ and we can see that as $A$ grows, the bigger the event horizon radius becomes. We should also note that the horizon $r_+$ is also a root of $h(r)$ even though $h(r)\neq b(r)$ (in fact $h(r)/b(r)=e^{-\frac{A^2}{r^2}}$), therefore, the black hole has the same causal structure as the static BTZ black hole, where inside the horizon %the $t, r$ coordinates shift and
we still have one time and two position coordinates. There exists a singularity at the origin as can be seen by calculating the Kretschmann scalar, which is plotted in Fig.~\ref{kre}. Its expression is complicated, but by checking the limits we can see that it is divergent at the origin, while regular for any other $r>0$ and at large distances is related to the cosmological constant
\begin{equation} R_{\alpha\beta\gamma\delta}R^{\alpha\beta\gamma\delta}(r\to \infty) \sim \frac{12}{\ell ^4}+\frac{8 A^2}{r^2 \ell ^4}+\mathcal{O}\left(\left(\frac{1}{r}\right)^4\right)~.
\end{equation}

\subsection{Energy Conditions}
\begin{figure}[h]
\centering
\includegraphics[width=.40\textwidth]{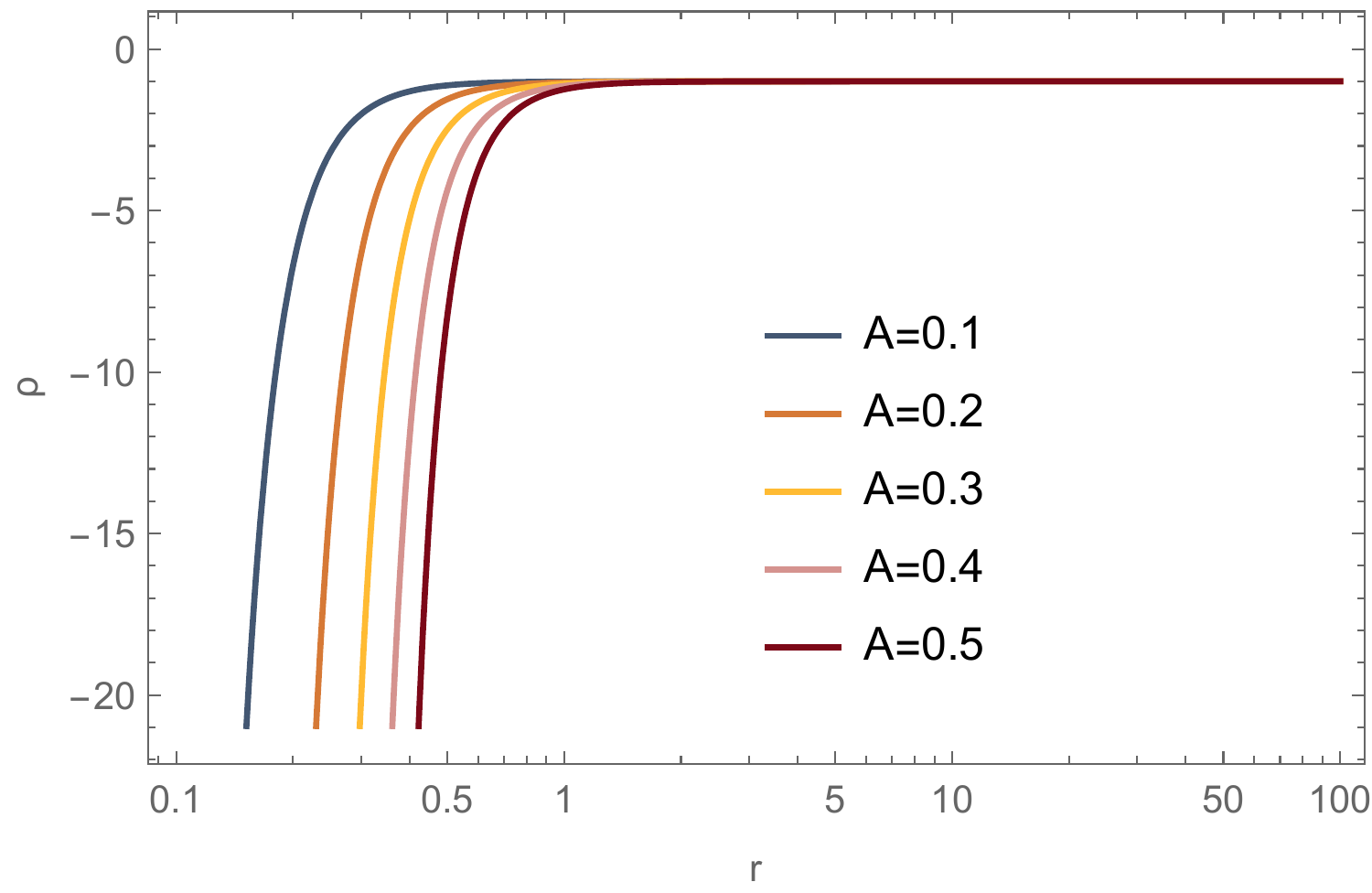}
\includegraphics[width=.39\textwidth]{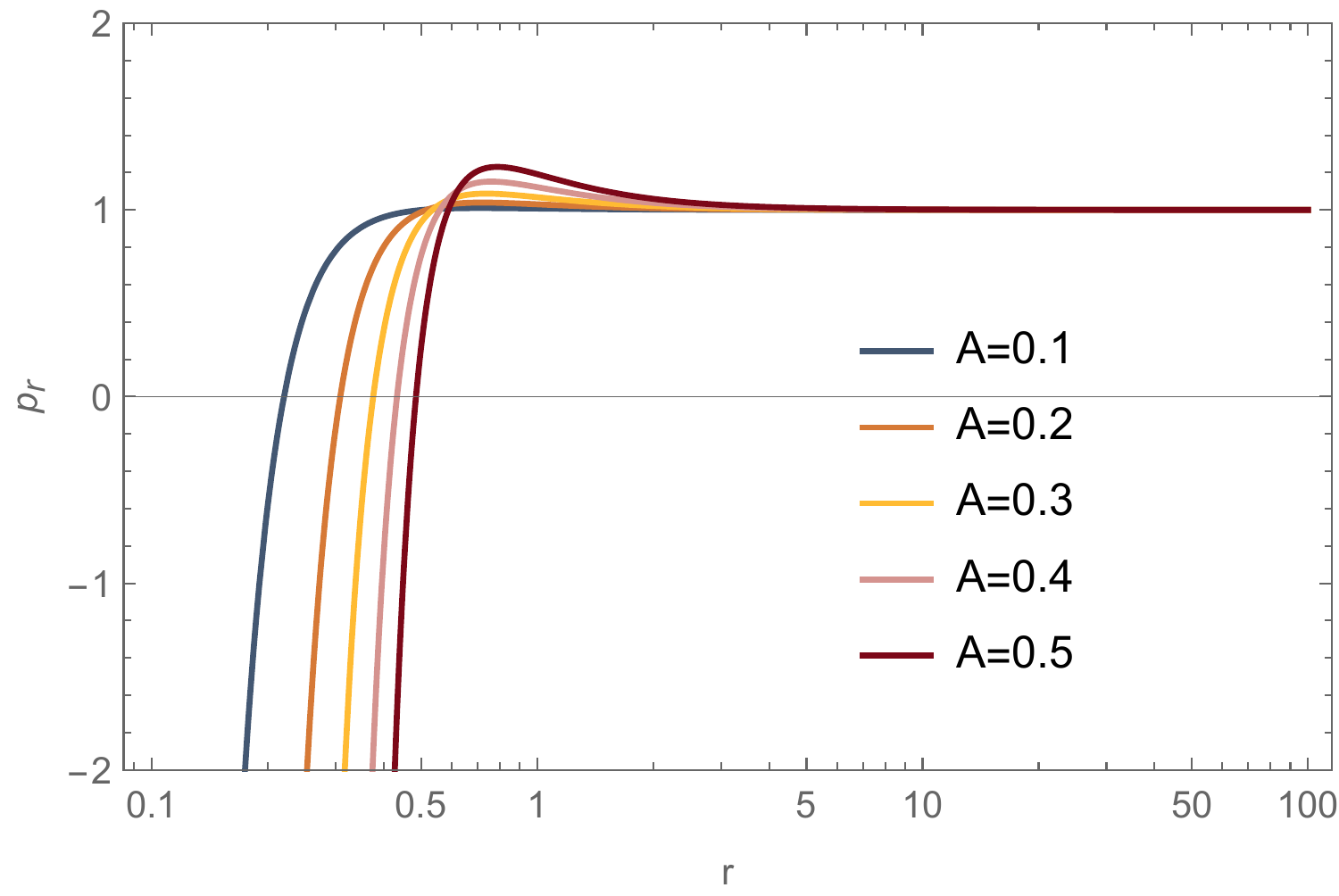}
\includegraphics[width=.40\textwidth]{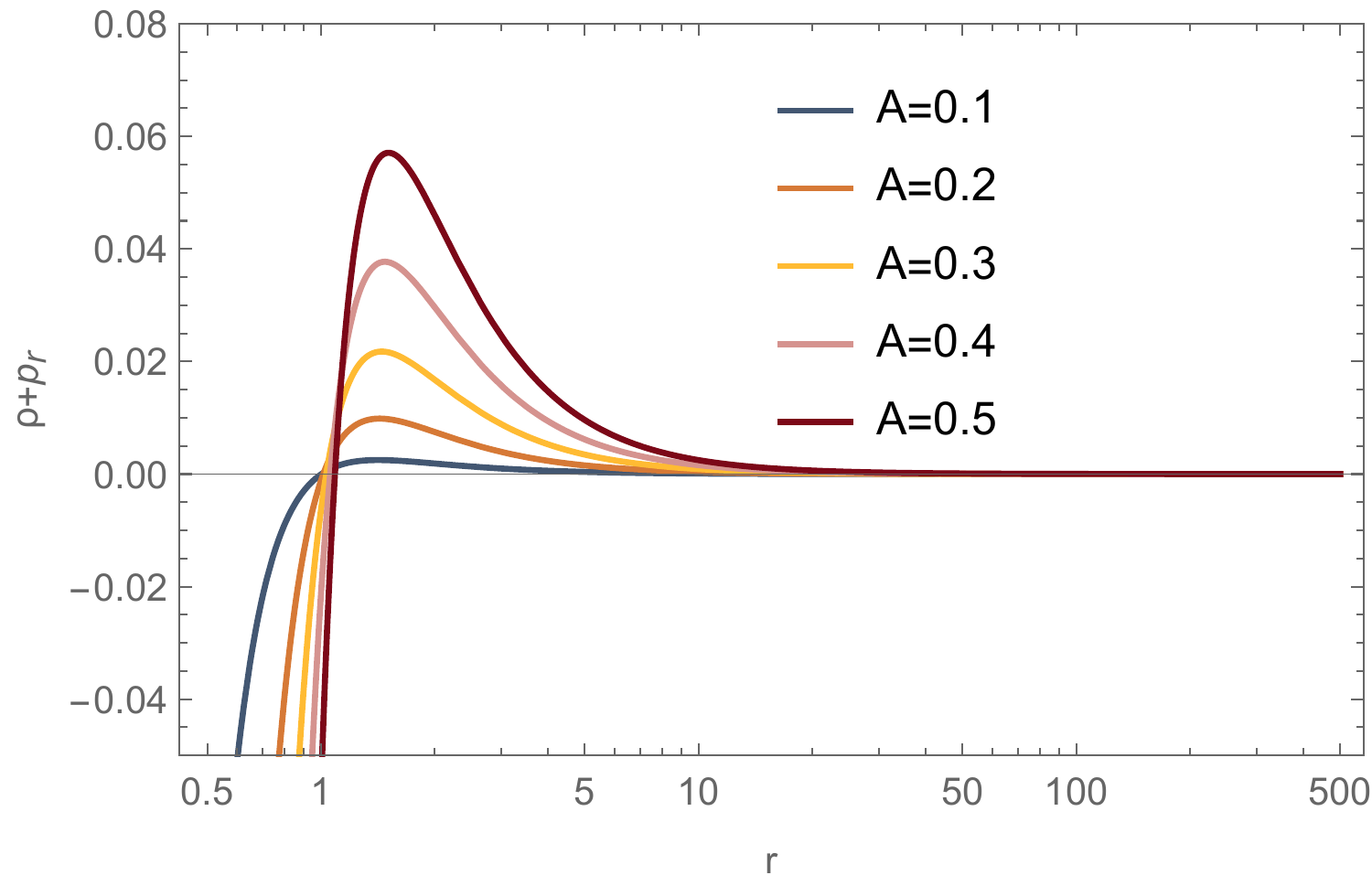}
\caption{The energy density $\rho$, the radial pressure $p_r$ and the radial pressure $\rho+p_r$ for $\mathcal{M}=\ell=1$ while changing the scalar charge $A$.} \label{ec}
\end{figure}

In this subsection we will discuss the energy conditions of the obtained space-time. For this reason we rewrite the Einstein field equation as
\begin{equation} G_{\mu}^{~\nu} = T_{\mu}^{~\nu}~.\end{equation}
In this frame of reference, we can identify $T_{t}^{~t} =-\rho,~ T_{r}^{~r}=p_r,~  T_{\theta}^{~\theta} = p_{t}=-\rho$ being the energy density, the radial pressure and the transverse pressure respectively. The energy conditions are obtained from these expressions. The weak energy condition (WEC) states that given a time-like vector field $t^{a}$, the quantity $T_{ab}t^at^b$ is positive, $T_{ab}t^at^b\ge0 \to \rho>0$. The null energy condition (NEC) states that $T_{ab}l^al^b\ge0 \to \rho + p_r>0$, where $l^al_a=0$,  so that the geometry will have a focussing effect on null geodesics. The explicit expressions read
\begin{eqnarray}
&& \rho =\mathcal{T}+ V = b(r)\phi'^2/2+V(r)  =\frac{e^{\frac{A^2}{2 r^2}} \left(A^2-2 r^2\right) \left(A^2+ \ell ^2
   \mathcal{M}\right)-e^{\frac{A^2}{r^2}} (A-r) (A+r) \left(A^2+2 \ell ^2 \mathcal{M}\right)}{A^2
   r^2 \ell ^2}~,\\
&& p_{r} = \mathcal{T}-V =b(r)\phi'^2/2-V(r) = \frac{e^{\frac{A^2}{2 r^2}} \left(A^2+2 r^2\right) \left(A^2+ \ell ^2 \mathcal{M}\right)-r^2
   e^{\frac{A^2}{r^2}} \left(A^2+2 \ell ^2 \mathcal{M}\right)}{A^2 r^2 \ell ^2}~,\\
&&\rho+p_{r} = 2\mathcal{T} =  b(r)\phi'(r)^2 = \frac{2 e^{\frac{A^2}{2 r^2}} \left(A^2+ \ell ^2 \mathcal{M}\right)-e^{\frac{A^2}{r^2}}
   \left(A^2+2 \ell ^2 \mathcal{M}\right)}{r^2 \ell ^2}~,
\end{eqnarray}
where $\mathcal{T}=b(r)\phi'^2/2$ is the kinetic energy of the scalar field. The energy density $\rho$ is negative inside and on the black hole horizon. Inside the horizon, $b(r)$ is negative, $\phi'^2$ is always positive and the potential is negative everywhere as we can see from Fig.~\ref{V}, resulting in negative energy density inside the black hole. On the horizon, we have $b(r_+)=0$, hence the contribution from the kinetic energy of the scalar field vanishes and the potential makes the energy density negative. Outside of the black hole horizon $b(r)$ is positive, but the effect of the potential energy is stronger than the kinetic energy, resulting in negative energy density everywhere, as we can see in Fig.~\ref{ec}. At large distances, the kinetic energy asymptotes as
\begin{equation}
\mathcal{T}(r\to\infty) \sim \frac{A^2}{2 r^2 \ell ^2}-\frac{A^2 \mathcal{M}}{ 2r^4}+\mathcal{O}\left(\left(\frac{1}{r}\right)^6\right)~,
\end{equation}
while the expression for the potential energy in the limit $r \to +\infty$ is given in (\ref{pt}). The leading order term in the kinetic energy is positive and since all constants $A,\ell, \mathcal{M}$ are finite constants, the kinetic energy is positive for large $r$. We can see in (\ref{pt}) that the potential will cancel this positive contribution of the kinetic energy, therefore the sum $\rho=\mathcal{T}+V$ will be always negative. It is known that too negative a potential might threaten the WEC of a regular scalar field. In our case the potential is the quantity that violates WEC.
For the NEC, we can see that at the event horizon $r_+$, we have $\rho +p_{r_+}=0$ due to the fact that $b(r_+)=0$. Outside and on the horizon the NEC is satisfied, while inside the event horizon the NEC is violated. This is a common feature of black hole space-times that arise from an action that consists of the Ricci scalar of Einstein's gravity and a simple non-minimally self interacting scalar field in arbitrary dimensions that satisfies $g^{rr}(r_+)=0, \phi = \phi(r)$, and not a peculiar case of our model. This behaviour is indeed present in the four-dimensional case \cite{Bakopoulos:2021dry,Bakopoulos:2023hkh}. In Fig.~ \ref{ec} we plot the energy density (WEC) and the sum of the energy density and radial pressure (NEC) of our black hole in order to illustrate the discussion above. In FIG. \ref{ec} we also plot the radial pressure. We can see that the radial pressure is negative for some region inside the black hole horizon, while at the horizon and outside of the horizon, the radial pressure is positive.

\subsection{Thermodynamics}

\begin{figure}[h]
\centering
\includegraphics[width=.40\textwidth]{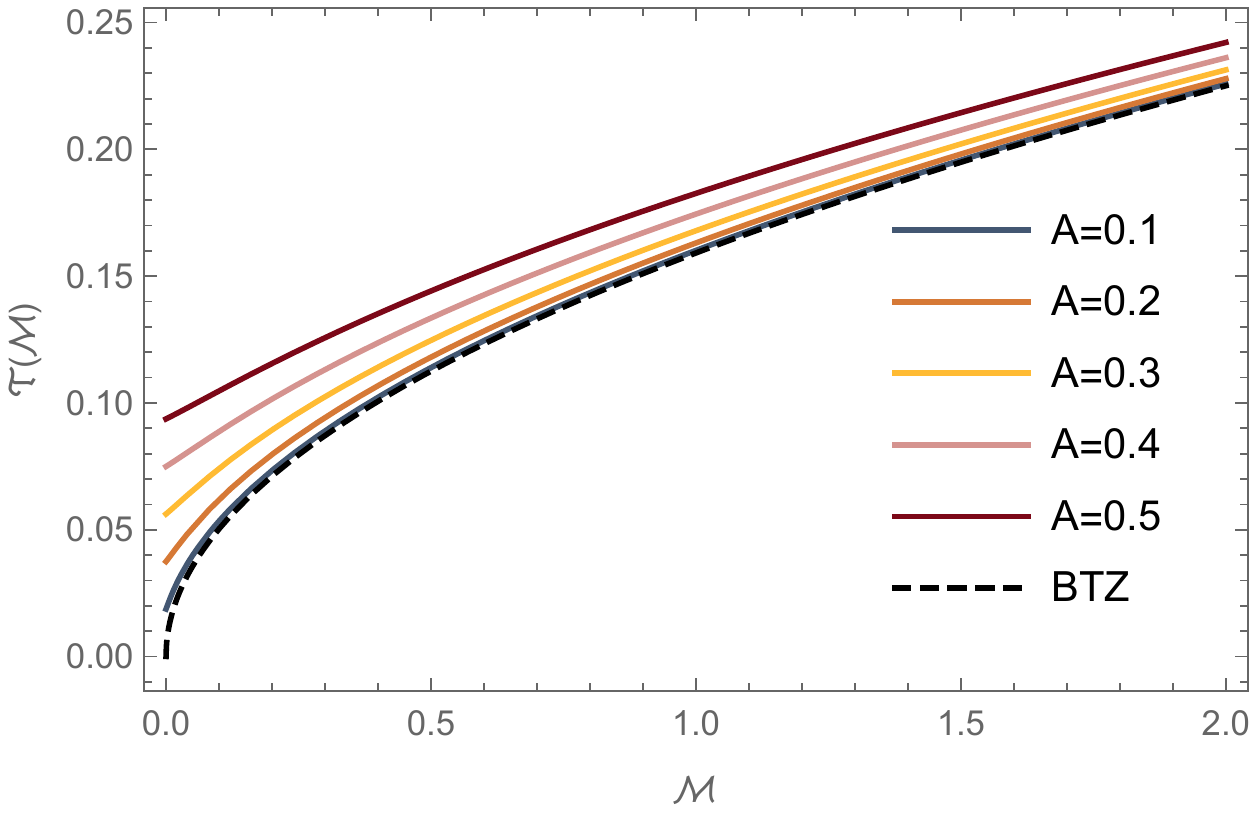}
\caption{The temperature of the black hole for $\ell=1$ as a function of $\mathcal{M}$, while changing the scalar charge.} \label{temp}
\end{figure}

In this subsection we will discuss the thermodynamics of the black hole solution. We begin with the temperature. To derive the black hole temperature, at first we perform a Wick rotation and move to imaginary time $t \to i\tau$ where $\tau$ will now be periodic, the period of which we have to find in order to specify the temperature. We will now ignore the angular part of the space-time metric and thence we are left with
\begin{equation} ds^2 = h(r)d\tau^2 + \frac{1}{b(r)}dr^2~.\end{equation}
We now expand the metric functions near the horizon
\begin{eqnarray}
&&h(r\to r_+) = h(r_+) + h'(r_+)(r-r_+) + ... = h'(r_+)(r-r_+)~,\\
&&b(r\to r_+) = b(r_+) + b'(r_+)(r-r_+) + ... = b'(r_+)(r-r_+)~,
\end{eqnarray}
and the reduced space-time element reads
\begin{equation} ds^2 = h'(r_+)(r-r_+)d\tau^2 + \frac{1}{b'(r_+)(r-r_+)}dr^2~.\end{equation}
Even if it is not clear at this point, the above line element describes a cone in Euclidean space and has a conical singularity at the tip $r\to 0$, unless we fix the period of $\tau$ in a particular way. Therefore, we will now compare this line element with the line element of two-dimensional flat space-time in polar coordinates that reads
\begin{equation} dS^2 = dR^2 +R^2d\Theta^2~,\end{equation}
where $\Theta$ is periodic of period $T_{\Theta} = 2\pi$ and we will treat $\tau$ as an angular coordinate, in order for the space-time to be truly Euclidean. By setting $ds^2=dS^2$ we can relate the two radial coordinates
\begin{equation} dR^2 = \frac{1}{b'(r_+)(r-r_+)}dr^2~,\end{equation}
which by integration will yield the relation
\begin{equation} R=2\sqrt{\frac{r-r_+}{b'(r_+)}}~,\end{equation}
and now we are left with the angular coordinates
\begin{equation} h'(r_+)(r-r_+)d\tau^2 = R^2d\Theta^2~,\end{equation}
which again by integration yields
\begin{equation} \Theta = \frac{\sqrt{h'(r_+)b'(r_+)}}{2}\tau \to \frac{ \Theta}{\tau} = \frac{\sqrt{h'(r_+)b'(r_+)}}{2}~.\end{equation}
$\Theta$ is periodic with $T_{\Theta} = 2\pi$ and by denoting $\beta$ the period of $\tau$ we have
\begin{equation}\beta = \frac{4\pi}{\sqrt{h'(r_+)b'(r_+)}} \to \mathfrak{T} \equiv \frac{1}{\beta} = \frac{\sqrt{h'(r_+)b'(r_+)}}{4\pi}~,\end{equation}
which is the temperature of our black hole space-time. Substituting the functions we find
\begin{equation}
\mathfrak{T}(r_+)=\frac{\left(A^2+2 r_+^2\right) \left(A^2+\ell ^2 \mathcal{M}\right)-r_+^2 e^{\frac{A^2}{2 r_+^2}} \left(A^2+2 \ell ^2 \mathcal{M}\right)}{2 \pi
   A^2 r_+ \ell ^2}~,
\end{equation}
and substituting the horizon radius we can express the temperature as a function of the black hole mass
\begin{equation} \mathfrak{T}({\mathcal{M}}) = \frac{\left(A^2+\ell ^2 \mathcal{M}\right) \sqrt{\ln \left(A^2+\ell ^2 \mathcal{M}\right)-\ln \left(A^2+2 \ell ^2 \mathcal{M}\right)+\ln
   (2)}}{\sqrt{2} \pi  A \ell ^2}~.
\end{equation}
The temperature is always real and positive. In Fig.~\ref{temp} we plot the temperature of the black hole.
We can see that the temperature increases as the mass of the black hole is growing. Moreover, the temperature is non-zero and finite for zero mass, which happens because a horizon exists even for the massless case, given by $r_+(\mathcal{M}=0) = A/(2\ln2)$. For small $A$, it becomes
\begin{equation} \mathfrak{T}(r_+) \sim \frac{r_+}{2 \pi  \ell ^2} + \frac{A^2 \left(2 r_+^2-\ell ^2 \mathcal{M}\right)}{8 \pi  r_+^3 \ell ^2}+\mathcal{O}\left(A^4\right)~,
\end{equation}
while the BTZ temperature corresponds to $\mathfrak{T}_{\text{BTZ}} =r_+/(2\pi\ell^2) = \sqrt{\mathcal{M}}/2 \pi  \ell $. As a result, the hairy black hole possesses a larger temperature at the event horizon.
Using the Wald  formula \cite{Wald:1993nt,Iyer:1994ys}, we can calculate the entropy of the black hole as
\begin{equation} \mathcal{S}= -2\pi\int d\theta \sqrt{r_+^2}\left(\frac{\partial \mathcal{L}}{\partial R_{\alpha\beta\gamma\delta}}\right)\Bigg|_{r=r_+}\hat{\varepsilon}_{\alpha\beta}\hat{\varepsilon}_{\gamma\delta}~, \end{equation}
where $\hat{\varepsilon}_{\alpha\beta}$ is the bi-normal to the horizon surface \cite{Dutta:2006vs}, $\mathcal{L}$ is the Lagrangian of the theory $\mathcal{L} =(8\pi)^{-1}\left( R/2  -\partial^{\alpha}\phi\partial_{\alpha}\phi/2 - V(\phi)\right)$, and
\begin{equation} \frac{\partial \mathcal{L}}{\partial R_{\alpha\beta\gamma\delta}}\Big|_{r=r_+} = \frac{1}{2}\left(g^{\alpha \gamma}g^{\beta\delta} - g^{\beta\gamma}g^{\alpha\delta}\right)~,\end{equation}
hence the entropy will be given by the Bekenstein-Hawking area law
\begin{equation} \mathcal{S} = \frac{\mathcal{A}}{4} = \frac{\pi r_+}{2}~,\label{entropy} \end{equation}
where $\mathcal{A} = 2\pi r_+$ is the circumference of the three-dimensional black hole. The hairy black holes possess a larger event horizon radius in comparison to the BTZ black hole which has an event horizon at $r_+=\ell \sqrt{\mathcal{M}}$
\begin{equation} r_+ \sim \ell  \sqrt{\mathcal{M}}+\frac{3 A^2}{8 \ell  \sqrt{\mathcal{M}}}+\mathcal{O}\left(A^3\right)~,\end{equation}
hence they are thermodynamically preferred over the BTZ black hole, having higher entropy, when the scalar charge is small. The black hole space-time is thermally stable in the canonical ensemble, which can be seen by evaluating the heat capacity
\begin{equation} C(r_+) = T\frac{dS}{dT}\Big|_{r=r_+}=\frac{\pi  A}{2 \sqrt{2} \sqrt{\ln \left(A^2+\ell ^2 \mathcal{M}\right)-\ln \left(A^2+2 \ell ^2 \mathcal{M}\right)+\ln (2)}}~,\end{equation}
which is always positive. For negligible scalar hair we obtain the heat capacity of the BTZ black hole $C = \pi r_+/2 = \pi\ell\sqrt{\mathcal{M}}/2$.

In general, one cannot impose that the internal energy of the black hole coincides with the mass of the black hole, as the matter fields may contribute to the internal energy. As a result the first law cannot simply be $d\mathcal{M} =T dS$ but $dE = TdS$, where $E$ is the internal energy. We can re-express the horizon of the black hole as
\begin{equation} r_+ = \frac{A}{\sqrt{\ln \left(\frac{4 \left(q \ell ^2+1\right)^2}{\left(2 q \ell ^2+1\right)^2}\right)}}~,
\end{equation}
where we have substituted the mass parameter with $q$ via (\ref{cond}). It is therefore clear, that the only free parameter of the system that one is allowed to vary is $A$. Now, we have that
\begin{equation} TdS = T\frac{\partial S}{\partial r_+}dr_+ =T\frac{\partial S}{\partial r_+}\frac{\partial r_+}{\partial A}dA = T\frac{\partial S}{\partial A}dA = \frac{A \left(q \ell ^2+1\right) \sqrt{\ln \left(2 q \ell ^2+2\right)-\ln \left(2 q \ell ^2+1\right)}}{2 \sqrt{2} \ell ^2 \sqrt{\ln \left(\frac{4 \left(q \ell ^2+1\right)^2}{\left(2 q \ell ^2+1\right)^2}\right)}}~.
\end{equation}
Integrating the above expression we get:
\begin{equation} E(A) = \frac{A^2 \left(q \ell ^2+1\right) \sqrt{\ln \left(2 q \ell ^2+2\right)-\ln \left(2 q \ell ^2+1\right)}}{4 \sqrt{2} \ell ^2 \sqrt{\ln \left(\frac{4 \left(q \ell ^2+1\right)^2}{\left(2 q \ell ^2+1\right)^2}\right)}}~,
\end{equation}
which is the internal energy of the black hole. It is clear that when the effect of the scalar field theory is negligible we have that
\begin{equation} E(q\to 0) \sim \frac{A^2 q}{8}+\frac{A^2}{8 \ell ^2} = \frac{\mathcal{M}_{\text{BTZ}}}{8} + \frac{A^2}{8 \ell ^2}~,
\end{equation}
which is a nice way to express that the fields contribute to the total energy of the black hole. We note that the conserved black hole mass and the geometric mass in the BTZ black hole differ by a factor of $1/8$.

\section{Rotating Black Hole Solutions} \label{sec3}

\begin{figure}[h]
\centering
\includegraphics[width=.40\textwidth]{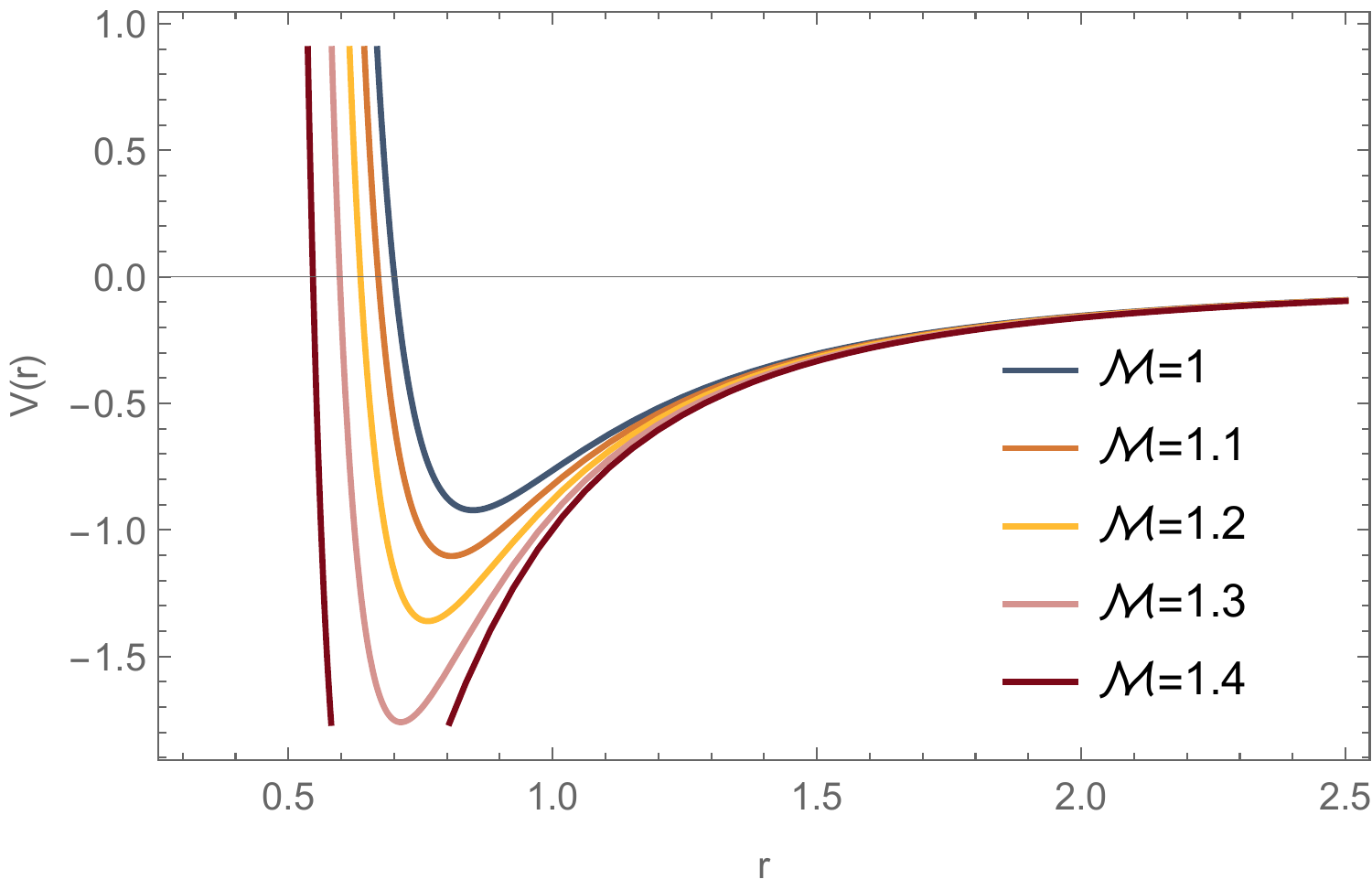}
\includegraphics[width=.40\textwidth]{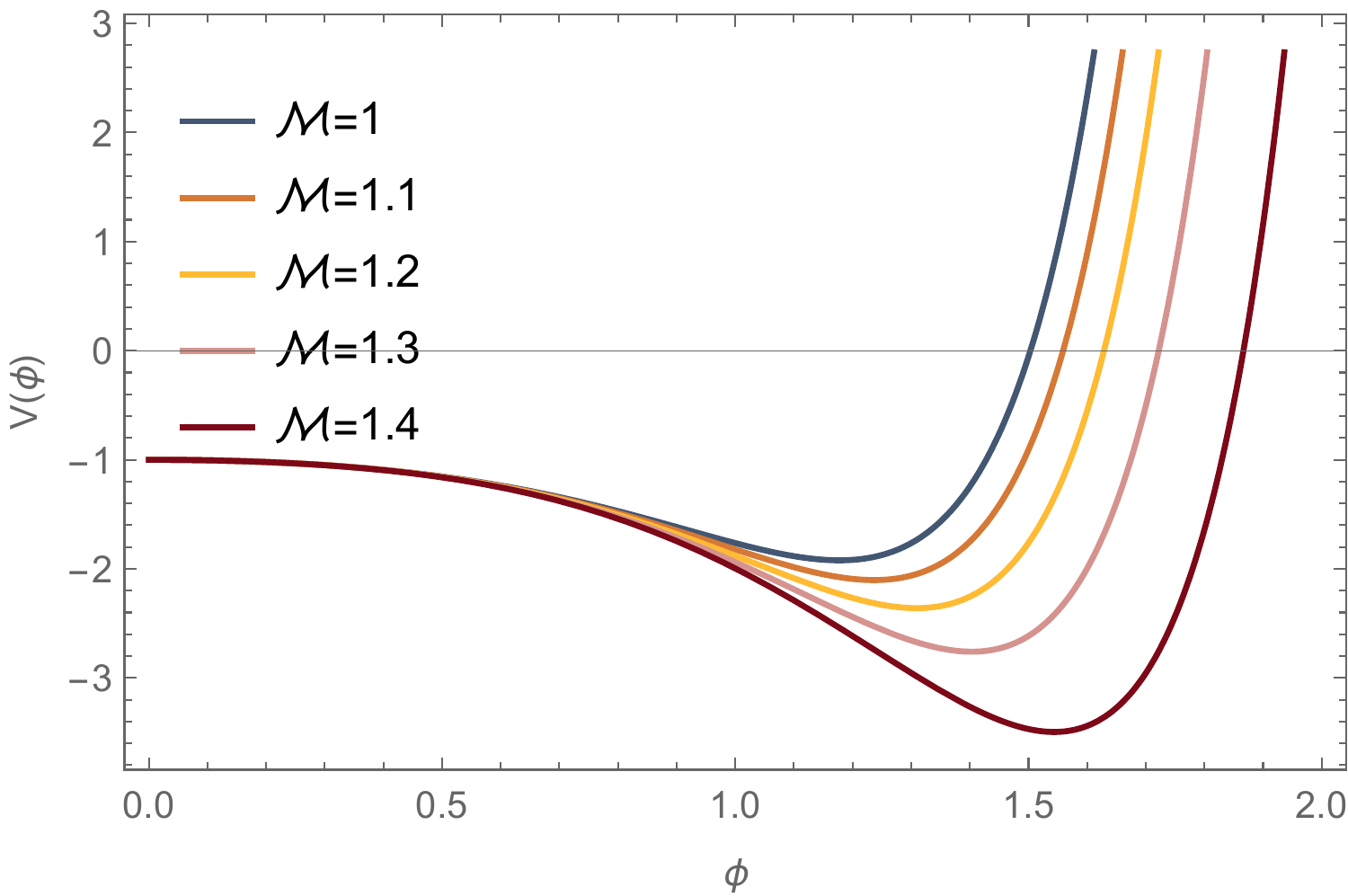}
\caption{$V(r)$ having set $\ell=A=1$, $J=2$, while varying the mass parameter.} \label{Vrot}
\end{figure}

\begin{figure}[h]
\centering
\includegraphics[width=.39\textwidth]{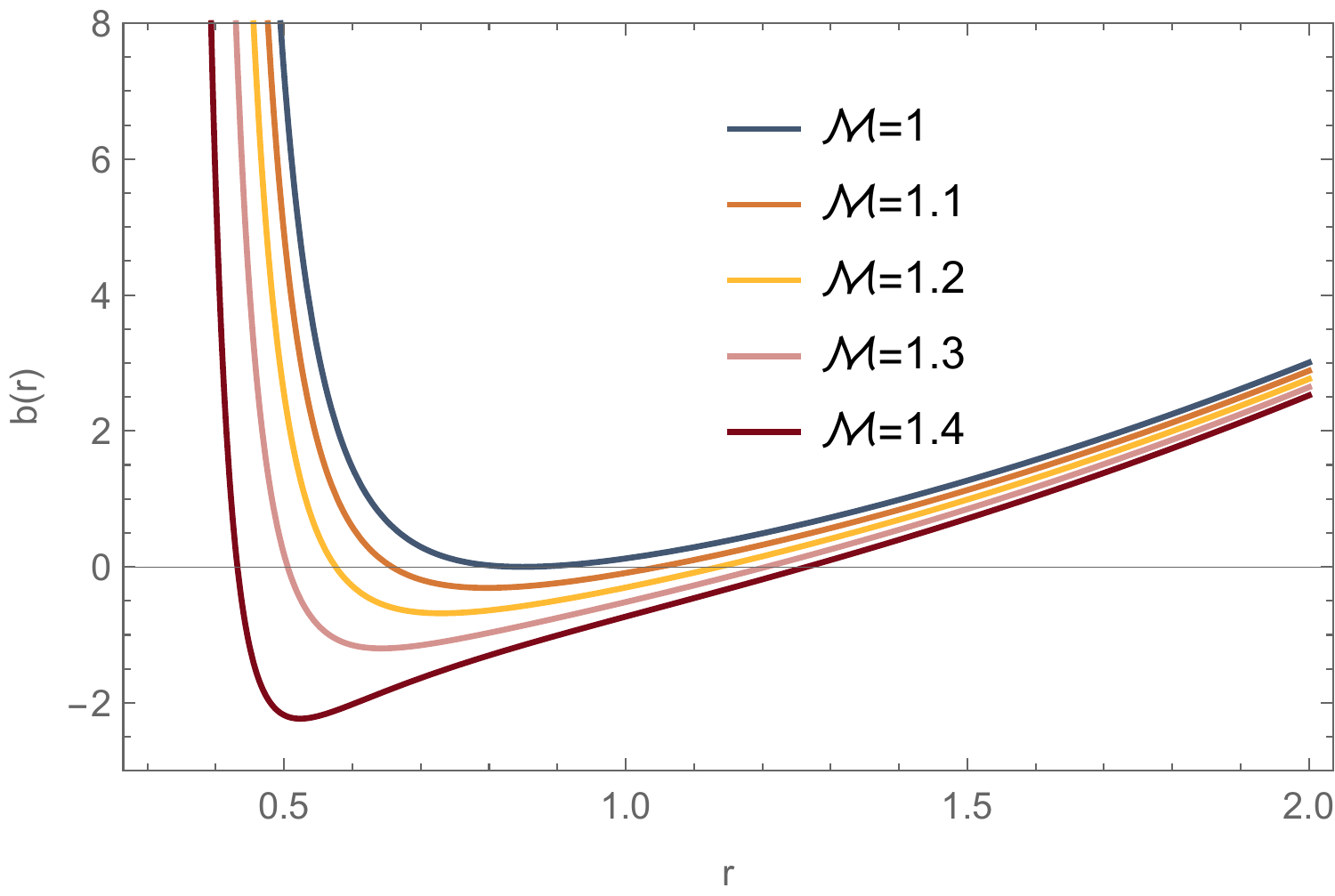}
\includegraphics[width=.40\textwidth]{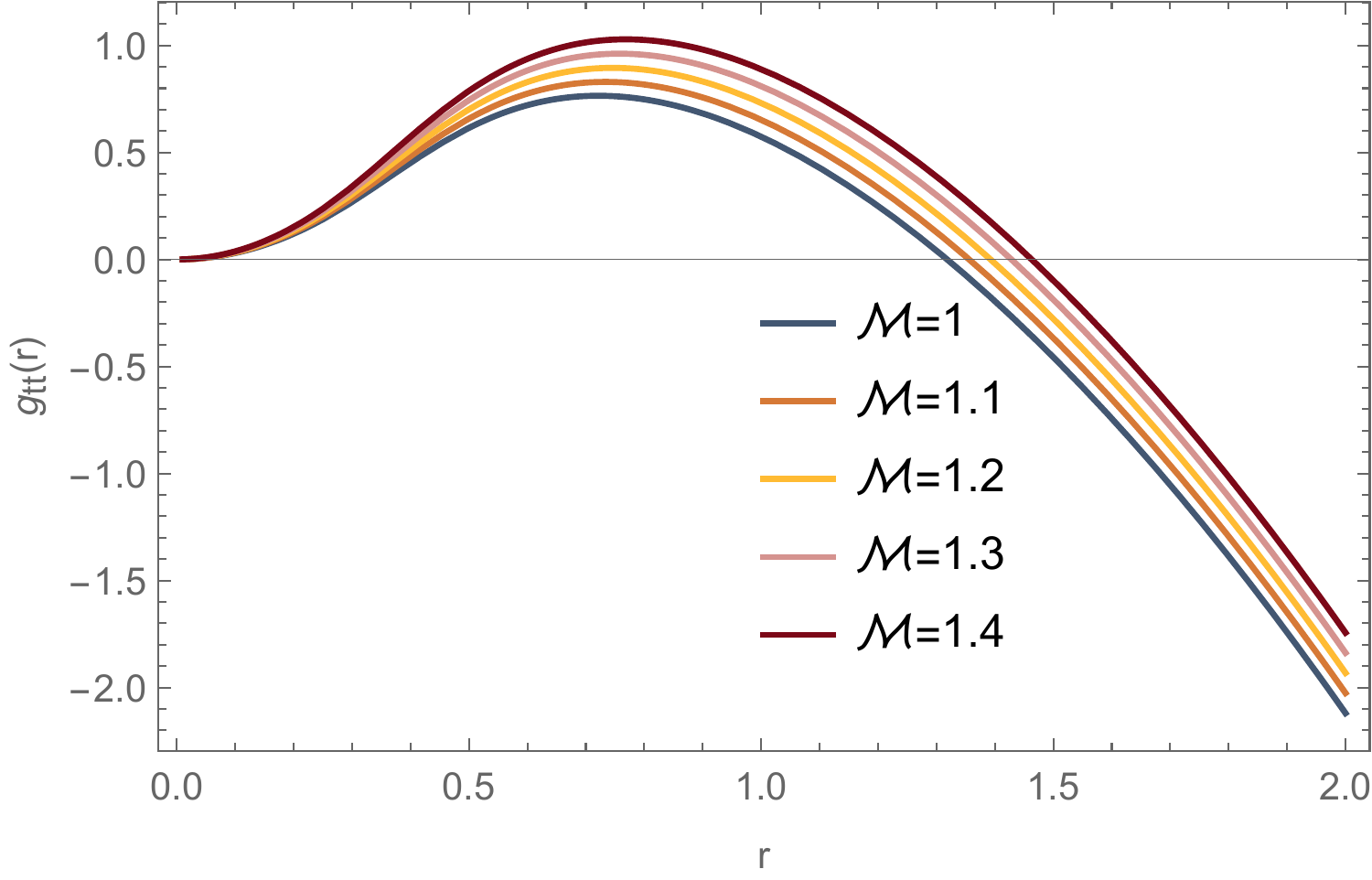}
\caption{$b(r)$ and $g_{tt}(r)$ having set $\ell=A=1$, $J=2$, while varying the mass parameter.} \label{rot}
\end{figure}

\begin{figure}[h]
\centering
\includegraphics[width=.40\textwidth]{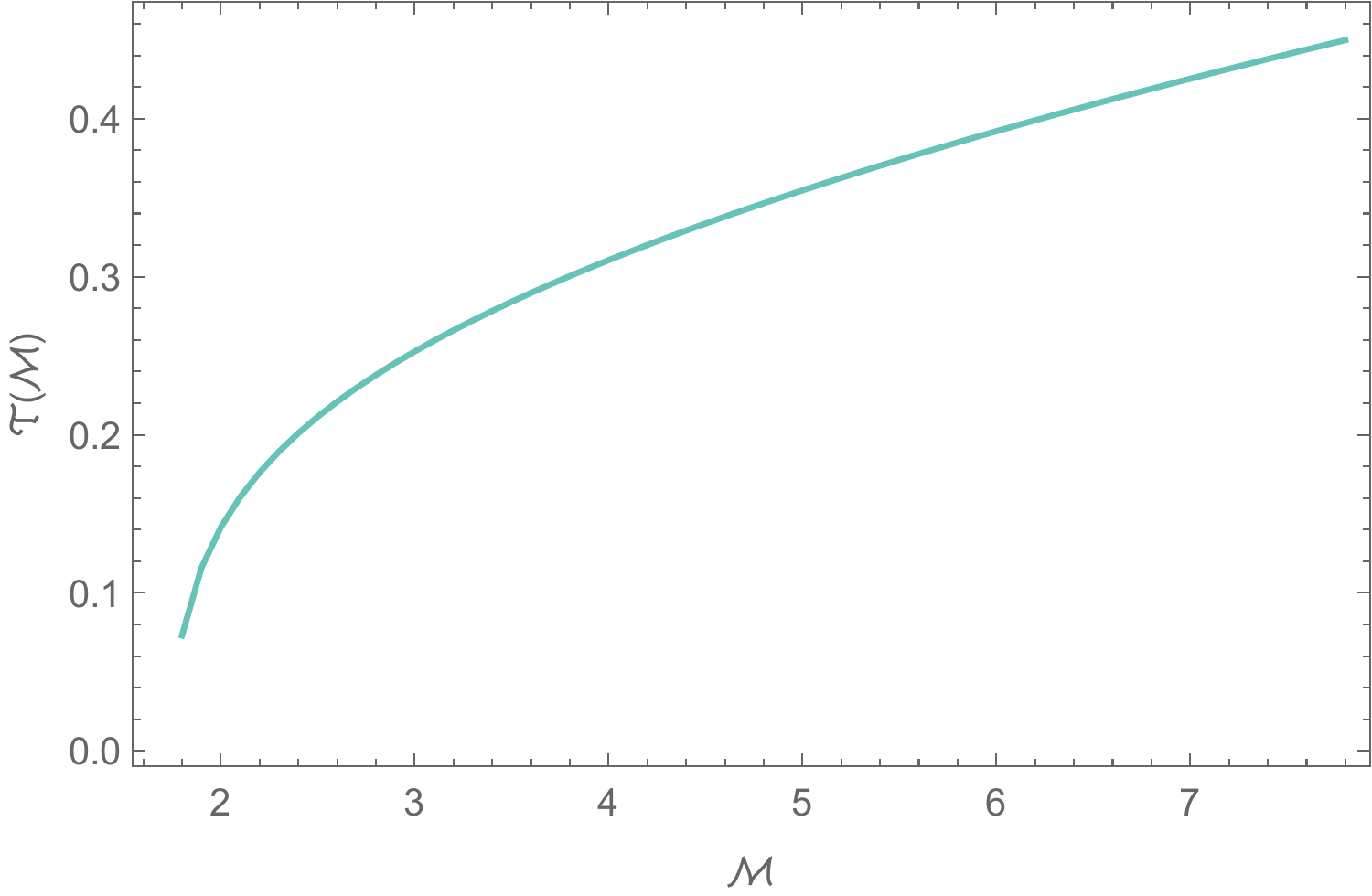}
\includegraphics[width=.39\textwidth]{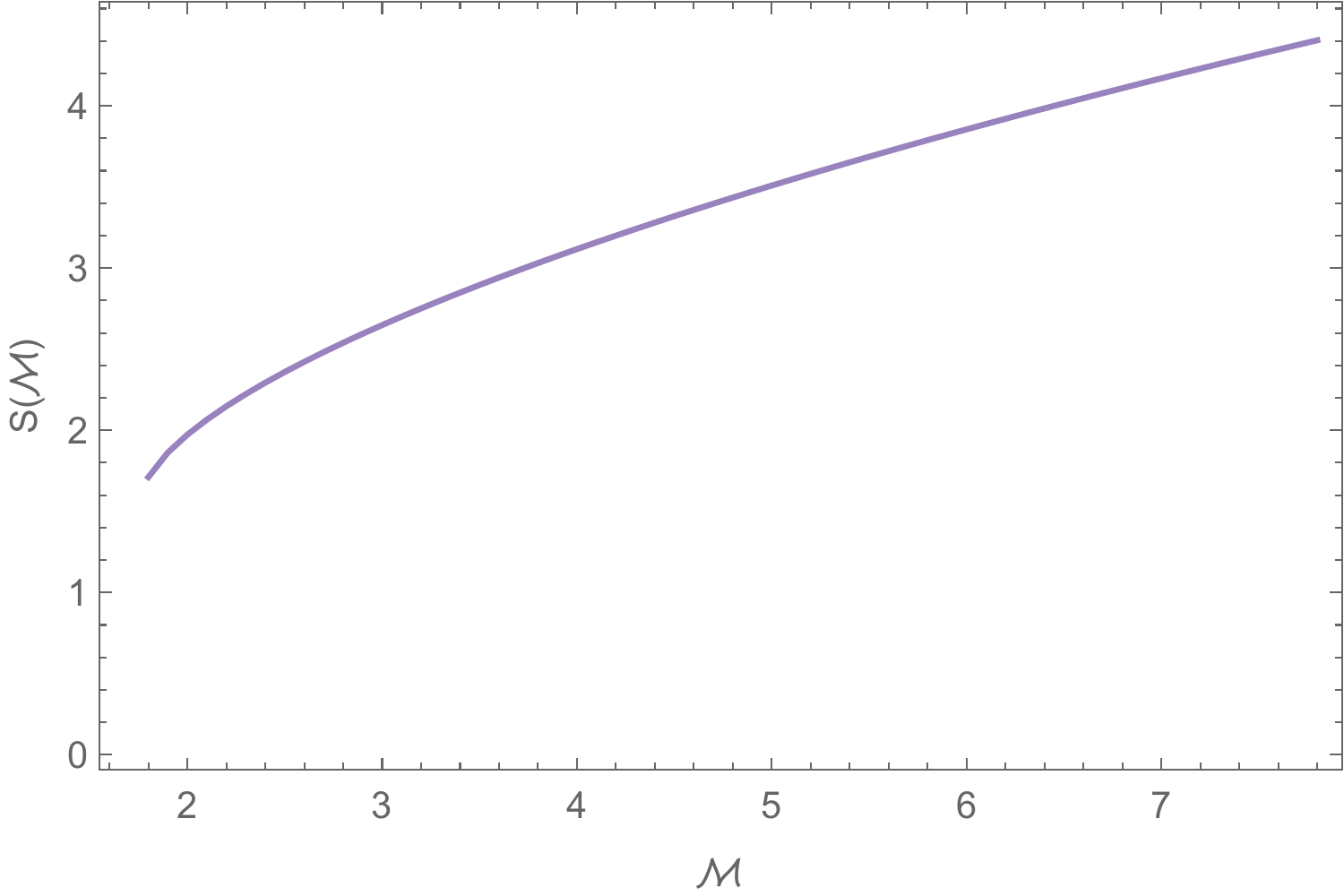}
\includegraphics[width=.40\textwidth]{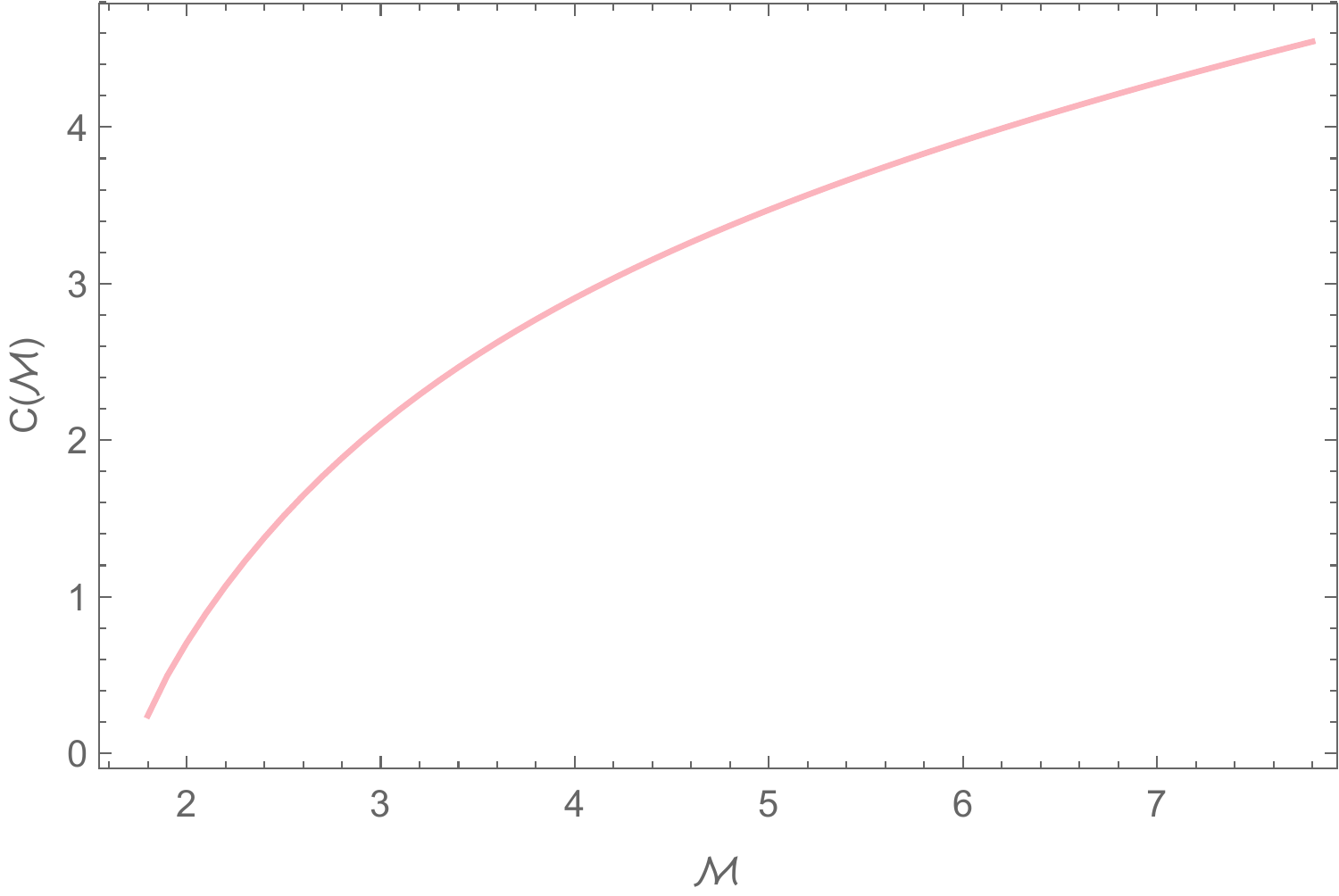}
\caption{The temperature $\mathfrak{T}$, the entropy $S$ and the heat capacity $C$ as functions of the black hole mass} \label{rotating_thermo}
\end{figure}

To discuss rotating solutions, we impose the metric ansatz
\begin{equation} ds^2 = -h(r)dt^2 + \frac{1}{b(r)}dr^2 + r^2\left(d\theta + u(r)dt\right)^2~,\end{equation}
where we have introduced the angular shift function $u(r)$. Inserting this ansatz in the field equations (\ref{einstein}),(\ref{box}) we obtain the following solution
\begin{eqnarray}
&& u(r) = \frac{J }{A^2}e^{-\frac{A^2}{2 r^2}}-\frac{J}{A^2}~,\\
&& b(r) = \frac{r^2 e^{\frac{A^2}{2 r^2}} \left(-\frac{2 J^2}{A^2}+\frac{2 A^2}{\ell ^2}+2
   \mathcal{M}\right)}{A^2}+\frac{r^2 e^{\frac{A^2}{r^2}} \left(-A^2 \left(-\frac{2
   J^2}{A^2}+\frac{2 A^2}{\ell ^2}+2 \mathcal{M}\right)+\frac{A^4}{\ell
   ^2}-J^2\right)}{A^4}+\frac{J^2 r^2}{A^4}~,\\
&& h(r) = e^{-A^2/r^2}b(r)~,\\
&&V(r) = -\frac{8 r^4 e^{\frac{A^2}{2 r^2}} \left(A^2 \ell ^2 \mathcal{M}+A^4-J^2 \ell ^2\right)+2 r^2
   e^{\frac{A^2}{r^2}} \left(A^2-2 r^2\right) \left(2 A^2 \ell ^2 \mathcal{M}+A^4-J^2 \ell
   ^2\right)+J^2 \ell ^2 \left(2 A^2 r^2+A^4+4 r^4\right)}{4 A^4 r^4 \ell ^2}~,\\
&&V(\phi) = -\frac{8 e^{\frac{\phi ^2}{2}} \left(A^2 \ell ^2 \mathcal{M}+A^4-J^2 \ell ^2\right)+2 e^{\phi ^2}
   \left(\phi ^2-2\right) \left(2 A^2 \ell ^2 \mathcal{M}+A^4-J^2 \ell ^2\right)+J^2 \ell ^2
   \left(\phi ^4+2 \phi ^2+4\right)}{4 A^4 \ell ^2}~,
\end{eqnarray}
while the scalar field is the same as (\ref{phi}) and $J$ is the angular momentum of the black hole. We used the quasi-local method to derive the angular momentum and the conserved black hole mass \cite{Chan:1995wj}
\begin{eqnarray}
&&J =\lim_{r_0\to\infty} \frac{\sqrt{b(r_0)}u'(r_0)r_0^3}{\sqrt{h(r_0)}}~,\\
&&\mathcal{M} = \lim_{r_0\to\infty}\left( \sqrt{h(r_0)}E(r_0) -Ju(r_0)\right)~.
\end{eqnarray}
 The asymptotic expressions at large distances yield
\begin{eqnarray}
&&u(r \to \infty) \sim -\frac{J}{2 r^2}+\frac{A^2 J}{8 r^4}-\frac{A^4 J}{48 r^6}+\frac{A^6 J}{384
   r^8}+\mathcal{O}\left(\left(\frac{1}{r}\right)^{10}\right)~,\\
&&h(r\to\infty) \sim \frac{r^2}{\ell ^2}+\left(-\frac{A^2}{\ell ^2}-\mathcal{M}\right)+\frac{\frac{A^4}{\ell ^2}+A^2
   \mathcal{M}+J^2}{4 r^2}-\frac{3 A^2 J^2 \ell ^2+A^4 \ell ^2 \mathcal{M}+A^6}{24 r^4 \ell
   ^2}+\mathcal{O}\left(\left(\frac{1}{r}\right)^6\right)~,\\
&&b(r\to\infty) \sim \frac{r^2}{\ell ^2}-\mathcal{M}+\frac{-\frac{A^4}{\ell ^2}-3 A^2 \mathcal{M}+J^2}{4 r^2}+\frac{A^2
   \left(-\frac{3 A^4}{\ell ^2}-7 A^2 \mathcal{M}+3 J^2\right)}{24
   r^4}+\mathcal{O}\left(\left(\frac{1}{r}\right)^6\right)~,\\
&&V(r\to \infty) \sim -\frac{1}{ \ell ^2}-\frac{A^2}{2 r^2 \ell ^2}-\frac{A^2 \ell ^2 \mathcal{M}+A^4}{4 r^4 \ell
   ^2}+\mathcal{O}\left(\left(\frac{1}{r}\right)^6\right)~,\\
&&V(\phi \to 0) \sim -\frac{1}{ \ell ^2}-\frac{\phi ^2}{2 \ell ^2}+\phi ^4 \left(-\frac{\mathcal{M}}{4 A^2}-\frac{1}{4 \ell
   ^2}\right)+\frac{1}{24} \phi ^6 \left(\frac{3 J^2}{A^4}-\frac{5 \mathcal{M}}{A^2}-\frac{3}{\ell
   ^2}\right)+\mathcal{O}\left(\phi ^8\right)~.
\end{eqnarray}
We can see that at large distances we obtain a solution similar to the rotating BTZ black hole, with changes in the structure of space-time being related to $A$, and in the small hair case, we also obtain the rotating BTZ black hole. The horizon is given by $b(r)=0$ \cite{Clement:1995zt,Hennigar:2020drx}, which is also a root of $h(r)$. $b(r)$ has two roots, which can be computed analytically but these expressions are lengthy, so we will not given them here. The existence of horizons provides bounds for the angular momentum of the black hole. Therefore, black holes can only exist when
\begin{equation}A>0~\& ~\ell >0~\&~ \mathcal{M}>0~\&~ \left(J^2\leq\frac{A^4}{\ell ^2}+2 A^2 \mathcal{M} \text{~or~} \frac{A^4}{\ell ^2}+2 A^2 \mathcal{M}<J^2\leq \frac{\left(A^2+\ell ^2 \mathcal{M}\right)^2}{\ell ^2}\right)~. \label{ineq} \end{equation}
In the cases where the inequalities are saturated ($J^2=A^4/\ell ^2+2 A^2 \mathcal{M}$ or $J^2 = \left(A^2+\ell ^2 \mathcal{M}\right)^2/\ell ^2$) in the above expressions we have black holes with a single event horizon, while in any other case the black holes develop two horizons, an inner and an event horizon.
The rotating solution admits a region of space-time where the Killing field $\partial/\partial t$ is space - like, which transforms into the condition
\begin{equation} g_{tt} >0 \to -h(r) +r^2u(r)^2>0~.\end{equation}
It is clear that at the event horizon, the previous condition holds, and also for some region outside of the horizon \cite{Hennigar:2020drx}.
In Fig.~\ref{Vrot} we plot the scalar potential where it is obvious that between the inner and event horizon, negative potential wells are developed, while as we increase the mass parameter, the wells become deeper. In addition, it is clear form both figures, that in the rotating case, the vaccum of the field theory represents a local maximum, and not a global one. One should also note the fact that the potential energy is bounded from below, which will play a crucial role in the stability of the system.  Moreover,  in Fig.~\ref{rot} we plot the metric function $b(r)$ and $g_{tt}$ in order to study the geometry.
From Fig.~\ref{rot} (left) we can see that the black hole develops two horizons, while for the degenerate case $J^2=\left(A^2+\ell ^2 \mathcal{M}\right)^2/\ell ^2$, the horizons coincide. The $g_{tt}$ component is positive at the horizon and for some region after the horizon, denoting the presence of an ergo-region, as in the BTZ black hole \cite{Banados:1992wn,Banados:1992gq,Carlip:1995qv}. The scalar potential depends on the conserved black hole charges. We can eliminate these charges by introducing a new constant $\chi$, such as
\begin{equation} \chi = \frac{J^2}{A^4}~,\end{equation}
besides $q=\mathcal{M}/A^2$. Now the potential will read
\begin{eqnarray}
&&V(\phi) =e^{\phi ^2} \left(\phi ^2 \left(-q+\frac{\chi }{2}-\frac{1}{2 \ell ^2}\right)+2 q-\chi
   +\frac{1}{\ell ^2}\right)+e^{\frac{\phi ^2}{2}} \left(-2 q+2 \chi -\frac{2}{\ell ^2}\right)-\chi
   -\frac{\chi  \phi ^4}{4}-\frac{\chi  \phi ^2}{2} ~,\\
&&V(\phi \to 0) \sim -\frac{1}{\ell ^2} -\frac{\phi ^2}{2 \ell ^2}+\frac{\phi ^4 \left(-q \ell ^2-1\right)}{4 \ell ^2}+\frac{\phi ^6 \left(-5
   q \ell ^2+3 \chi  \ell ^2-3\right)}{24 \ell ^2}+\mathcal{O}\left(\phi ^8\right)~.
\end{eqnarray}
Consequently, our theory will give black hole space-times with a fixed angular momentum to conserved mass ratio given by
\begin{equation} \frac{\chi}{q^2} = \frac{J^2}{\mathcal{M}^2}~.\end{equation}
Both the conserved mass and the angular momentum are allowed to vary, in a consistent way, so the ratio $J^2/\mathcal{M}^2$ will always be constant.
To compute the Hawking temperature we will use the concept of surface gravity, which is defined by
\begin{equation} \kappa = \sqrt{-\frac{1}{2}\nabla_{\mu}X_{\nu}\nabla^{\mu}X^{\nu}}~,\end{equation}
where $X^{\mu}$ is a Killing vector field of our space-time defined as $X^{\mu} = (1,0,\Omega)$, where $\Omega$ is the angular velocity at the horizon of the black hole, defined as
\begin{equation} \Omega = -\frac{g_{t\theta}}{g_{\theta\theta}}\bigg|_{r=r+}~.\end{equation}
Evaluating the surface gravity we find
\begin{equation} \kappa = \frac{1}{2} \sqrt{-\frac{b(r) \left(h(r) \left(r u'(r)+2 (u(r)+\Omega )\right)^2-\left(h'(r)-r^2
   (u(r)+\Omega ) u'(r)\right)^2\right)}{h(r)}}\Bigg|_{r=r+}~.\end{equation}
Now, we  can find the temperature as
\begin{equation} \mathfrak{T}(r_+) =\frac{\kappa}{2\pi} = \frac{1}{4 \pi }\sqrt{b(r) \left(\frac{h'(r)^2}{h(r)}-r^2 u'(r)^2\right)}\Bigg|_{r=r+} = \frac{A^4 e^{\frac{A^2}{r_+^2}}-J^2 \ell ^2 \left(e^{\frac{A^2}{2 r_+^2}}-1\right){}^2}{4 \pi  A^2
   r_+ \ell ^2 \left(e^{\frac{A^2}{r_+^2}}-e^{\frac{A^2}{2 r_+^2}}\right)}~,\end{equation}
which of course reduces to the temperature of the BTZ black hole, when $A$ approaches $0$.
The entropy is given by the same formula as in the non-rotating case (\ref{entropy}).
The heat capacity is found to be
\begin{equation} C(r_+) =\pi  r_+^3\left(2 \left(A^2 \left(J \ell  \left(\frac{1}{J \ell -e^{\frac{A^2}{2 r_+^2}}
   \left(A^2+J \ell \right)}+\frac{1}{e^{\frac{A^2}{2 r_+^2}} \left(A^2-J \ell \right)+J \ell
   }\right)-3\right)+r_+^2\right)\right)^{-1}~,
\end{equation}
which for small $A$ reduces to the heat capacity of the rotating BTZ black hole
\begin{equation} C(r_+,A\to0) = \frac{1}{6} \pi  r_+ \left(\frac{16 r_+^4}{3 J^2 \ell ^2+4 r_+^4}-1\right)>0~.\end{equation}
Due to the complexity of (\ref{ineq}) in order for the existence of horizons, we will perform a numerical analysis for the thermodynamics of this solution and plot the temperature, the entropy, and the heat capacity as functions of the black hole mass $\mathcal{M}$ for appropriate values of $A,J,\ell$. In Fig.~\ref{rotating_thermo} we fix $A=0.5,\ell=1,J=2$ and plot the corresponding thermodynamic quantities for the allowed values of mass.
It is clear all thermodynamic quantities are positive and growing with the increase of mass, in accordance with the BTZ black hole case. Moreover, these black holes are thermally stable in the canonical ensemble and do not develop any phase transition.  The system under consideration is complicated and therefore we cannot present simple calculations to show that the internal energy of the black hole do not coincide, however we do not expect a different behaviour from the non-rotating case.

\section{Conclusions} \label{sec4}

In this work we consider a three-dimensional Gravity Theory with a self-interacting scalar field minimally coupled to Gravity. We do not specify the form of the scalar potential and to solve the field equations we choose a specific form of the scalar field which is characterized by a scalar charge. Solving the field equations we find a solution in which the scalar field back - reacts to the metric and a hairy black hole is generated. The presence of the scalar charge is introducing a scale in the theory and in the scalar potential an effective cosmological constant appears. Motivated by field theory, we introduce a parameter as the ratio of the black hole mass to the scalar charge and this parameter appears in the potential. In this way the potential is independent of the mass of the black hole and the scalar charge. The introduction of this parameter allows  the black hole mass  to vary, along with the scalar charge in a particular way, so that their ratio is kept constant.

We find that the form of the hairy black hole depends on the way matter parametrized by the scalar field back - reacts to the background gravitational metric. At large distances because matter is weekly back - reacting to Gravity, the BTZ black hole is generated with the scalar charge playing the role of the cosmological constant which is necessary for the existence of the BTZ black hole. At small distances a novel hairy BTZ-like black hole is produced,  because matter is strongly back - reacting to the gravitational metric. We discussed the thermodynamics of the solutions, where we found that the hairy black hole possesses a larger event horizon than the BTZ case and hence has a higher temperature and entropy at the event horizon. We also discuss the first law of the thermodynamics which for the hairy BTZ-like black hole has to be modified.

Finally we discuss rotating solutions introducing angular momentum through an angular shift function. We find a similar behaviour as in the non-rotating case, at large distances we obtain a solution similar to the rotating BTZ black hole, with changes in the structure of space-time being related to the scalar charge, while in the small scalar charge case, we  obtain the rotating BTZ black hole. At small distances the rotating BTZ black hole is modified but because of the complexity we did not give an explicit form of the rotating hairy BTZ-like black hole at small distances. Performing a numerical analysis for the thermodynamics of this solution and studying the temperature, the entropy, and the heat capacity as functions of the black hole mass, the scalar charge and the angular momentum we found  a similar behaviour for the non-rotating case of the hairy BTZ-like black hole.

\section{Acknowledgements}
We thank Athanasios Bakopoulos,  Christos Charmousis, Alex Kehagias, Nick E. Mavromatos and Theodoros Nakas for useful discussions.
The research project was supported by the Hellenic Foundation for Research and Innovation (H.F.R.I.) under the “3rd Call for H.F.R.I. Research Projects to support Post-Doctoral Researchers” (Project Number: 7212). E.P and G.K. are also supported by the research project of the National Technical University of Athens (NTUA)
65232600-ACT-MTG, entitled: Alleviating Cosmological Tensions Through Modified Theories of Gravity.

\end{document}